\documentstyle[preprint,aps]{revtex}
\tightenlines
\input psfig.sty
\begin{document}
\title { Proton Structure Functions from Chiral Dynamics and QCD Constraints}
\author{H. J. Weber }
\address{ Institute of Nuclear and Particle Physics, University of
Virginia, \\Charlottesville, VA 22903, USA}
\maketitle
\begin{abstract}
The spin fractions and deep inelastic lepton structure functions of 
the proton are analyzed using chiral field theory involving Goldstone 
bosons. A detailed comparison with recent chiral models sheds light on  
their successful description of the spin fractions of the proton as being due 
to neglecting helicity non-flip chiral transitions. This approximation is  
valid for zero mass but not for constituent quarks. Since the chiral spin 
fraction models with the pure spin-flip approximation reproduce the measured 
spin fractions of the proton, axialvector constituent-quark-Goldstone boson 
couplings are found to be inconsistent with the proton spin data. Initial 
quark valence distributions are then constructed using quark counting 
constraints at Bjorken $x\rightarrow$1 and Regge behavior at $x\rightarrow0$. 
Sea quark distributions predicted by chiral field theory on this basis have 
correct order of magnitude and shape. The spin fractions also agree with the 
data.   
\end{abstract}
\vskip0.5in
\par
PACS numbers: 11.30.Rd,\ 12.39.Fe,\ 14.20.Dh
\par
Keywords: Deep inelastic scattering, chiral field theory, QCD constraints    
\newpage

\section{Introduction }

The nonrelativistic quark model (NQM) explains qualitatively many of 
the strong, electromagnetic and weak properties of the nucleon and other octet 
 (and decuplet) baryons in terms of three valence quarks whose dynamics is 
motivated by quantum chromodynamics (QCD), the gauge field theory of the 
strong interaction.
The effective degrees of freedom at the scale $\Lambda _{QCD}$ are dressed or 
dynamical quarks along with Goldstone bosons which are expected to emerge in 
the spontaneous chiral symmetry breakdown ($\chi SB$) of QCD that is 
characterized by quark condensates. While the NQM 
description of baryon states has not been derived from first principles of QCD,
solutions of Schwinger-Dyson equations for light quarks with models for color  
confinement lead to a momentum dependent quark mass $m_q(p^2)$. Such dynamical 
quarks become constituent quarks when $m_q(p^2)$ is approximated by a   
constant $m_q(0)$ at the scale $\Lambda _{QCD}$. In Nambu--Jona-Lasinio 
models~\cite{HK} of the $\chi SB$ patterned after superconductivity and driven 
by a zero-range scalar and pseudoscalar quark-quark interaction, the 
cut-off $\Lambda _{NJL}\sim$ 631 MeV of the effective chiral field theory 
provides an estimate for the momentum scale where the dynamical quark mass 
from the gap equation has fallen to the current quark mass. A more realistic 
treatment of color confinement is provided by the dual superconductor picture 
in the 't Hooft abelian gauge~\cite{H} where QCD monopoles generated by the 
abelian gauge fixing condense in the QCD vacuum confining quarks by the dual 
Meissner effect. Here the dynamical $\chi SB$ leads to a quark mass consistent 
with the QCD running whose monopole shape parameter 
$\Lambda _m\sim \Lambda _{QCD}\sqrt{10}\sim$ 630 MeV is close to 
$\Lambda _{NJL}$~\cite{UST}. At this momentum scale current quarks become the 
relevant degrees of freedom.         
\par
At momentum scale $\Lambda _{QCD}$ chiral perturbation 
theory ($\chi$PT~\cite{CPT}) allows incorporating systematically the chiral 
dynamics of QCD. Chiral field theory applies 
to scales from $\Lambda _{QCD}$ up to a (presumed) chiral symmetry restoration 
scale $\Lambda _{\chi }=4\pi f_\pi = 1169$ MeV where $f_\pi = 93$ MeV is the 
pion decay constant. In this scenario, inside hadrons at distances that are 
smaller than the confinement scale $\Lambda ^{-1}_{QCD}$, but larger than 
distances where perturbative QCD (pQCD) applies, quarks and Goldstone bosons 
are the effective degrees of freedom. In principle, chiral field theory 
dissolves a dynamical or constituent quark into a current quark and a cloud of 
virtual Goldstone bosons, but here only one layer of Goldstone bosons is 
included. Thus, the effective degrees of freedom and interactions of 
chiral field theory are consistent with those of chiral perturbation theory.   
\par
Chiral fluctuations of valence quarks inside hadrons, 
$q_{\uparrow,\downarrow}\rightarrow q_{\downarrow,\uparrow}+(q\bar q')_0$, 
into pseudoscalar mesons, $(q\bar q')_0$, of the SU(3) flavor octet of 
Goldstone bosons, were first applied to the spin problem of the proton in 
ref.~\cite{EHQ}. It was shown that chiral dynamics can help one understand not 
only the reduction of the proton spin carried by the valence quarks from 
$\Delta \Sigma =1$ in the NQM to the experimental value of about $1/3$, but 
also the reduction of the axial vector coupling constant $g_A^{(3)}$ from the 
NQM value 5/3 to about 5/4. It is well known that relativistic effects reduce 
the axial charge further, and this causes problems for the spin fractions in 
relativistic quark models~\cite{WB} forcing the proton size parameter to a 
rather large value of about 1 fm. This problem will no longer occur when we 
replace the NQM by constraints from the SU(6) spin-flavor wave function of the 
proton. In addition, the violation of the Gottfried sum 
rule~\cite{Go} $GSR=1/3+2(\bar u-\bar d)/3$, which signals an isospin 
asymmetric quark sea in the proton, became plausible. The SU(3) symmetric 
chiral quark model explains several spin and sea quark observables of the 
proton, but not all of them. The data~\cite{E143,SMC} call for SU(3) breaking 
because some of the spin fractions such as the ratio of triplet to octet spin 
fractions $\Delta_3/\Delta_8=$5/3 and the weak axial vector coupling constant 
of the nucleon, $g_A^{(3)}={\cal F}+{\cal D}$,  
disagree with experiments in the SU(3) symmetric case. In~\cite{WSK,SMW} 
the effects of SU(3) breaking were more systematically built into these chiral 
models and shown to lead to a remarkable further improvement of the spin and 
quark sea observables (that are integrated over Bjorken x) in comparison with 
the data. It was also shown~\cite{WSK} that the $\eta '$ meson proposed 
in~\cite{CLi} gives an almost negligible contribution to the spin fractions of 
the nucleon at the present level of experimental errors not only because of 
its large mass but also due to the small singlet chiral 
coupling constant. It is therefore ignored in the following, and this is 
consistent with the understanding that, due to the axial anomaly, the $\eta '$ 
meson is not a genuine Goldstone boson.

A major objective here is to construct proton structure functions so
that, when 
integrated over Bjorken x, they reproduce the recent successful chiral field 
theory results for the spin and flavor fractions of the proton~\cite{WSK,WB}. 
This goal can be achieved, but only upon {\bf neglecting helicity non-flip  
chiral transitions in the polarized splitting functions}.  The latter are also 
neglected in the unpolarized splitting functions, but this approximation is 
not nearly as critical. 
\par
Standard quark models like the nonrelativistic NQM or light-cone quark models 
are at scale $\Lambda _{QCD}$, and it is not clear how these models should be 
evolved to the scale $\Lambda _{\chi }$ in the non-perturbative regime where 
initial quark distributions are needed.   
In \cite{W} ratios of structure functions were found to be fairly insensitive 
to the uncertainties inherent in the evolution and were therefore recommended 
as a tool for testing quark models for missing physics. An example is the 
connection between the negative slope of the ratio of unpolarized structure 
functions, $F^{n}_{2}/F^{p}_{2}$, and a spin dependent attraction between the 
up and down quarks in the proton. In order to construct initial valence 
distributions, we turn to a more successful alternative which has recently 
been developed in \cite{BBS}, called BBS henceforth. This approach is based 
on quark counting rules or, more precisely, it implements the leading 
power-law behavior at $x\rightarrow 1$ of the helicity-dependent distributions 
that is known from the minimally connected perturbative diagrams in 
conjunction with constraints from the SU(6) proton spin-flavor wave function 
of the NQM. The initial parton 
distributions provided by BBS have been found surprisingly successful when 
evolved to the scale of the deep inelastic scattering (DIS) data from about 2 
GeV~\cite{LSS}. Therefore, this approach is adopted and used as a benchmark to 
assess chiral dynamics for its ability to generate intrinsic quark-antiquark 
excitations from initial valence quark distributions of the proton. We shall 
see that chiral field theory then predicts reasonable sea quark distributions. 
\par
The chiral field theory of parton distributions is described in Section II. 
The antiquark and polarized quark distributions are presented and discussed in 
Section III. Moments are given in Section IV and used in Section V to impose 
quark counting rules from pQCD on the valence part of the initial parton 
distributions. 

\section{Quark distributions from chiral field theory}
\par
If the spontaneous chiral symmetry breakdown in the infrared regime of QCD is 
governed by chiral $SU(3)_L \times SU(3)_R$ 
transformations, then the effective interaction between the octet of Goldstone 
boson fields $\Phi_i$ involves the axial-vector coupling    
\begin{eqnarray}
{\cal L}_{int}=-{g_A\over 2f_\pi }\sum_{i=1}^{8}\bar q \partial_\mu 
               \gamma^\mu \gamma_5 \lambda_i \Phi_i q         
\label{lint}
\end{eqnarray}
which can flip the polarization of quarks: 
$q_{\downarrow}\rightarrow q_{\uparrow}+GB$, etc. In Eq.~\ref{lint}, the 
$\lambda_i$, $(i=1,2,...,8)$ are Gell-Mann's SU(3) flavor 
matrices, and $g_A$ is the dimensionless axial vector quark coupling constant 
that is taken to be 1 here, while    
\begin{equation} 
g_A^{(3)}=\Delta u -\Delta d=\Delta_3={\cal F}+{\cal D}
         =(G_A/G_V)_{n\rightarrow p}, 
\label{gan}
\end{equation}
is the isotriplet axial vector coupling constant of the weak decay of the 
neutron, and $\Delta u$, $\Delta d$ and $\Delta s$ stand for the fraction of 
proton spin carried by the u, d and s quarks, respectively. They are defined 
by the matrix elements of the singlet, triplet, octet axial vector currents, 
$A^{(i)}_{\mu}$ for i$=$0,3,8 of the nucleon state at zero momentum 
transfer. It is also common to define the hypercharge or octet spin 
fraction $\Delta_8$ and the total proton spin 
$2S_z=\Delta \Sigma$ in the infinite momentum frame as  
\begin{equation}
\Delta_8=\Delta u +\Delta d -2\Delta s=3{\cal F}-{\cal D}, \qquad 
\Delta \Sigma=\Delta u +\Delta d +\Delta s.  
\label{deli}
\end{equation}
\par
The spin fractions that we discussed so far are related to the first moments 
of the polarized structure functions $g_{1}^{p}(x), g_{1}^{n}(x)$ of the 
proton which have been measured in a series of SLAC (E142, E143, 
E154~\cite{E143}) and CERN (EMC, NMC, SMC~\cite{SMC}) deep inelastic 
lepton scattering experiments with increasingly sophisticated analyses 
based on pQCD.    

\par
Returning to chiral field theory that we use for a description of DIS 
structure functions of the proton, note that, despite the nonperturbative 
nature of the chiral symmetry breakdown, the interaction between quarks and 
Goldstone bosons is small enough for a {\bf perturbative expansion} to apply. 
\par
Writing only the flavor dependence of the chiral interaction the SU(3) 
symmetric Eq.~\ref{lint} has the matrix form   
\begin{eqnarray}   
L_{int}= {g_A\over 2f_{\pi }}\sum_{i=1}^{8} \bar q \lambda_i \Phi_i q ,
\label{fint}
\end{eqnarray}
\begin{eqnarray}
 {1\over \sqrt 2}\sum_{i=1}^{8}\lambda_i \Phi_i =  \left( \begin{array}{c}  
{ 1\over \sqrt 2} \pi^0 + { 1\over \sqrt 6} \eta \qquad \pi^+ \qquad  K^+ \cr
  \pi^- \quad -{ 1\over \sqrt 2}\pi^0+{ 1\over \sqrt 6}\eta \quad K^0 \cr
   K^- \qquad \bar K^0 \qquad  -{2\over \sqrt 6}\eta  
\end{array}\right) .  
\label{flama}  
\end{eqnarray}
\par
{}From Eq.~\ref{lint} the following probability densities  
$f(u_{\uparrow} \rightarrow \pi^+ + d_{\downarrow})$,...
for chiral fluctuations of quarks can be organized as coefficients in chiral 
reactions: 
\begin{eqnarray}\nonumber 
u_{\uparrow} \rightarrow f_{u\rightarrow \pi ^{+}d}(x_{\pi }, 
{\bf k}^2_{\perp}) (\pi^+ + d_{\downarrow})
+f_{u\rightarrow \eta u} {1\over 6}(\eta +u_{\downarrow})
+f_{u\rightarrow \pi ^0 u} {1\over 2}(\pi^0 + u_{\downarrow})\\\nonumber
+f_{u\rightarrow K^{+}s} (K^+ + s_{\downarrow}),\\\nonumber
d_{\uparrow} \rightarrow  f_{d\rightarrow \pi ^{-} u} (\pi^- + u_{\downarrow})
+f_{d\rightarrow \eta d} {1\over 6}(\eta +d_{\downarrow})
+f_{d\rightarrow \pi ^0 d} {1\over 2}(\pi^0 + d_{\downarrow})\\\nonumber
+f_{d\rightarrow K^0 s} (K^0 + s_{\downarrow}),\\
s_{\uparrow} \rightarrow f_{s\rightarrow \eta s} {2\over 3}(\eta 
   +s_{\downarrow})+f_{s\rightarrow K^- u} (K^- + u_{\downarrow}) 
+f_{s\rightarrow \bar K^0 d} (\bar K^0 + d_{\downarrow}),  
\label{fluc}
\end{eqnarray}
and corresponding ones for the other initial quark helicity. After integrating 
over transverse momentum in the infinite momentum frame, the coefficients 
in Eq.~\ref{fluc} become the polarized (minus sign) and unpolarized (plus 
sign) chiral splitting functions, 
\begin{eqnarray}
P^{\pm}_{GB/q}(x) = \int d^2{\bf k}_{\perp}
f^{\pm}_{q\rightarrow GBq'}(x,{\bf k}_{\perp})
\end{eqnarray}
that determine the probability 
for finding a Goldstone boson of mass $m_{GB}$ carrying the longitudinal 
momentum fraction $x_{GB}$ of the parent quark's momentum and a recoil quark 
$q'$ with momentum fraction $1-x_{GB}$ for each fluctuation of Eq.~\ref{fluc}. 
\par   
In (non-renormalizable) chiral field theory with 
cut-off $\Lambda _{\chi}$ of ref.~\cite{EHQ}, the unpolarized chiral splitting 
functions are 
\begin{eqnarray} 
P^{+}_{q\rightarrow q'+GB}(x_{GB})={g_{A}^2\over f_{\pi}^2} {x_{GB}
(m_q+m_{q'})^2\over 32\pi ^2}\int^{t_{min}}_{-\Lambda _{\chi}^2} dt {\left[
(m_q-m_{q'})^2-t\right]\over (t-m_{GB}^2)^2},
\label{EHQ}         
\end{eqnarray}
where $t=-\left[{\bf k}_{\perp}^2+x_{GB}[m_{q'}^2-(1-x_{GB})m_q^2]\right]/
(1-x_{GB})$ is the square of the Goldsone boson four-momentum. Next we obtain 
the polarized splitting functions from the unpolarized ones using  
that the latter contain the sum of helicity non-flip and flip probabilities, 
while the former contain the difference of these chiral probabilities.  
Since quarks are on their mass shell in the light-front dynamics used
here, the axial vector quark-Goldstone boson interaction is equivalent 
to the simpler $\gamma _5$ interaction. Except for a common overall
factor, the relevant unpolarized chiral transition probability is
proportional to 
\begin{eqnarray}
-{1\over 2} tr[(\gamma \cdot p+m_q)\gamma_5(\gamma \cdot
p'+m'_q)\gamma_5]=2(pp'-m_q m'_q)=(m_q-m'_q)^2-k^2,\ 
\end{eqnarray}
where $2pp'=m'^2_q+m^2_q-k^2.$ This is the numerator in Eq.~\ref{EHQ}
which can also be written as
\begin{equation}  
{1\over 1-x_{GB}}[{\bf k}_{\perp}^2+[m_{q'}-(1-x_{GB})m_q]^2]
\label{nu}
\end{equation}
 and has the following physical interpretation. Recall that the axial-vector 
quark-Goldstone boson coupling $\gamma _{\mu }\gamma _5 k^{\mu }$ in 
Eq.~\ref{lint} involves the spin raising and lowering operators 
$\sigma _1\pm i\sigma _2$ in a scalar 
product with the transverse momentum ${\bf k}_{\perp}$ of the recoil quark,  
which suggests that the ${\bf k}^{2}_{\perp}$ term in Eq.~\ref{nu} represents 
the helicity-flip probability of the chiral fluctuation, while the 
longitudinal and time components $\gamma _5\gamma _{\pm}$ induce the 
non-spinflip probability which depends on the quark masses. This can be seen 
from the helicity non-flip probability 
\begin{eqnarray}
|\bar u'_{\uparrow}\gamma_5 u_{\uparrow}|^2
=|\bar u'_{\downarrow}\gamma_5 u_{\downarrow}|^2
\sim (m'_q-x' m_q)^2,\ x'=1-x_{GB}, 
\label{fli}
\end{eqnarray}
using light cone spinors and suppressing the spinor normalizations. Thus 
Eq.~\ref{fli} identifies the mass term in Eq.~\ref{nu} as the helicity 
non-flip chiral transition.  Similarly, the helicity-flip probability is 
obtained from
\begin{eqnarray}
|\bar u'_{\downarrow}\gamma_5 u_{\uparrow}|^2
\sim {\bf p'}_{\perp}^2+x'^2{\bf p}_{\perp}^2-x'{\bf p'}_{\perp}\cdot
{\bf p}_{\perp}
\end{eqnarray}
which, in frames where ${\bf p}_{\perp}=0,$ reduces to ${\bf k}_{\perp}^2,$
i.e. the net helicity flip probability generated by
the chiral splitting process. In the nonrelativistic limit, where 
$|{\bf p'}_{\perp}|<<m_{q'},|{\bf p}_{\perp}|<<m_q,$ clearly non-spinflip
dominates over spinflip, while spinflip dominates at high momentum. 
\par  
The polarized splitting function $P^-$ therefore has the same quark mass 
dependence as $P^+$, but involves the helicity flip probability with the 
opposite sign, i.e. has 
$${1\over 1-x_{GB}}[-{\bf k}_{\perp}^2+[m_{q'}-(1-x_{GB})m_q]^2]$$ 
in its numerator, which agrees with ref.\cite{SW}.   
\par 
The $q\rightarrow q'GB$ splitting functions are expected to satisfy number and 
momentum sum rules~\cite{Meln}. For example, if the quark $q$ is accompanied 
by two spectator quarks of a nucleon which splits (virtually) into a Goldstone 
boson and an octet baryon $B$ containing the recoil quark $q'$ along with the 
spectator quark pair, then the splitting function in Eq.~\ref{splf} becomes the 
hadronic $f_{GB/N}(x_{GB})$ for $N\rightarrow B+GB$, which equals 
$f_{B/N}(1-x_{GB})$, since the recoil quark $q'$ has longitudinal momentum 
fraction $1-x_{GB}$. Therefore, there are equal numbers of Goldstone bosons 
emitted by the nucleon, $\langle n\rangle_{GB/N}$, and virtual baryons 
accompanying them, $\langle n\rangle_{B/N}$. This symmetry is violated by 
Eq.~\ref{EHQ} of the model~\cite{EHQ}. Furthermore, momentum conservation 
holds, viz. $\langle x_{GB}\rangle_{GB/N}+\langle x_B\rangle_{B/N}
=\langle n\rangle_{GB/N}$. In order to satisfy the symmetry and momentum 
conservation laws, we follow ref.~\cite{Meln} and replace the cut-off in 
Eq.~\ref{EHQ} by a vertex function G in the splitting function integrand   
\begin{eqnarray}
f^{\pm}_{q\rightarrow GB q'}(x_{GB}, {\bf k}^2_{\perp})
={g^2_{GB qq'}\over 16\pi ^3}{|G_{GB qq'}(x_{GB},
{\bf k}^2_{\perp})|^2\over x_{GB}(1-x_{GB})}{((1-x_{GB})m_q-m_{q'})^2\pm
{\bf k}^2_{\perp}\over (1-x_{GB})(m_q^2-M^2_{GB q'})^2}, 
\label{splf}
\end{eqnarray}
where ${\bf k}_{\perp}$ is the transverse momentum of the recoil quark 
$q'$ whose longitudinal momentum fraction of the parent quark's is $1-x_{GB}$. 
Note that $$f^{\pm}_{u\rightarrow u\pi ^0}=f^{\pm}_{u\rightarrow d\pi ^+}
=f^{\pm}_{d\rightarrow d\pi ^0}=f^{\pm}_{d\rightarrow u\pi ^-}.$$ 
The vertex function 
$G_{GB qq'}$ in Eq.~\ref{splf} takes into account the extended structure of 
the Goldstone boson-quark system. It is taken here to depend on the inverse of 
the same covariant propagator $(m_q^2-M^2_{GB q'})^{-1}$ that occurs in the 
splitting function in Eq.~\ref{splf},   
\begin{eqnarray}
G_{GB qq'}(x_{GB},{\bf k}^2_{\perp})=
exp\left({m_q^2-M^2_{GB q'}(x_{GB},{\bf k}^2_{\perp})\over 2\Lambda ^2}\right)
\label{fzff}
\end{eqnarray}
with the invariant mass squared of the quark-Goldstone boson system $GB+q'$ 
\begin{eqnarray}
M^2_{GB q'}(x_{GB},{\bf k}^2_{\perp})
={m_{GB}^2+{\bf k}^2_{\perp}\over x_{GB}}
+{m^2_{q'}+{\bf k}^2_{\perp}\over 1-x_{GB}}. 
\label{MGBq'}
\end{eqnarray}
Such models satisfy the symmetry constraint mentioned above. 
\par 
The parameter $\Lambda $ in Eq.~\ref{fzff} 
controls the size of the Goldstone boson-quark system so that 
$2/\Lambda \sim <r^2>_{GB q'}^{1/2}\leq {1\over 3}$ fm. If we relate 
the monopole shape of $m_q(p^2)$ mentioned in the Introduction to a 
Gaussian shape, then $\Lambda \sim 2\sqrt{5}\Lambda _{QCD}/0.78\sim$1.14 GeV   
is obtained. Such hadronic descriptions of meson clouds of baryons are 
currently studied by many groups. Since the vertex-times-propagator form of 
the splitting functions in Eq.~\ref{splf} admits a wave function 
interpretation for the virtual hadronic process $N\rightarrow B+GB$, a 
light-front wave-function parametrization has been adopted in ref.~\cite{BMa} 
for $N\rightarrow \Lambda +K$.
\par 
The pion-quark coupling constant in Eq.~\ref{splf} may be obtained from 
the Goldberger-Treiman relation (GTR) 
\begin{eqnarray}
g_{\pi NN} = g_{A}^{(3)} {m_N\over f_{\pi }},
\label{GT}
\end{eqnarray}
where $f_{\pi }\sim 93$ MeV is the pion decay constant, in conjunction with   
the NQM coupling constant relation 
\begin{equation}
g_{\pi qq'}={3\over 5}g_{\pi NN}. 
\label{GTQ}
\end{equation}  
If a constituent quark mass instead of the nucleon mass is used in a GTR 
for constituent quarks as in Eq.~\ref{EHQ},
\begin{eqnarray}
g_{\pi qq'}=g_A{m_q+m_{q'}\over 2f_{\pi}}, 
\end{eqnarray}  
which is common in chiral quark models at scale $\Lambda _{QCD}$, then the 
factor $3/5$ in Eq.~\ref{GTQ} is replaced by $g_A/3g_{A}^{(3)}$, where $g_A$ is 
the quark axial-vector coupling constant often taken to be $3/4$ or 1.     
Since the factor $3/5$ in Eq.~\ref{GTQ} is scale (and model) dependent, we set 
$g_{\pi qq'}=\alpha _{\chi }g_{A}^{(3)}m_N/f_{\pi }$ and vary the chiral 
strength parameter $\alpha _{\chi }$ 
within the limits ${1\over 5}\leq\alpha _{\chi }\leq {3\over 5}$. In the 
SU(3) symmetric case $f_K\sim f_{\pi }$ for the kaons, and we set 
$g_{K qq'}=g_{\pi qq'}$. 

\par 
The integral over Bjorken x of the splitting function 
\begin{eqnarray}
P_{GB/q}=|\alpha _{GB/Q}|^2=\int_0^1 dx\int d^2 {\bf k}_{\perp}
f^{+}_{q\rightarrow GB q'},
\end{eqnarray}
is the probability of finding a virtual Goldstone boson $GB$ in the initial 
quark $q$. This in turn is related to the Fockspace expansion of the dressed 
quarks  (in light front field theory where it is well defined), e.g. 
\begin{eqnarray}
|U\rangle=\sqrt{Z}|u_0\rangle +\alpha _{\pi /U}|d\pi ^+\rangle
+{1\over 2}\alpha _{\pi /U}|u\pi ^0\rangle
+{1\over 6}\alpha _{\eta /U}|u\eta \rangle 
+\alpha _{K/U}|sK^+\rangle+... ,
\label{fock}
\end{eqnarray}
where the factors $p_{\pi ^0}=1/2$ for the neutral pion, $p_{\eta }=1/6$ and 
$p_{\pi ^{\pm}}=1=p_{K^{\pm}}=1$ come from the flavor dependence in 
Eq.~\ref{flama},  
$Z=1-P_u$ is the wave function renormalization constant   
and all masses and coupling constants are renormalized quantities.
\par
From the u and d quark lines in Eq.~\ref{fluc} the total meson emission 
probability $P_u$ of the u quark is given to first 
order in the Goldstone fluctuations by 
\begin{eqnarray}
P_u = \int^{1}_{0}dx\int d^2{\bf k}_{\perp}\left[{1\over 2}
f^{+}_{u\rightarrow u\pi ^0}+{1\over 6}f^{+}_{u\rightarrow u\eta }
+f^{+}_{u\rightarrow d\pi ^+}+f^{+}_{u\rightarrow sK^+}\right]
=\sum_{q,m}p_m \int^{1}_{0}dx P^{+}_{u\rightarrow q m},
\label{qprob}
\end{eqnarray}  
while the total strange quark probability density  
\begin{eqnarray}
P_s = \int^{1}_{0}dx\int d^2{\bf k}_{\perp}\left[{2\over 3}f^{+}_{s\rightarrow 
s\eta }+f^{+}_{s\rightarrow u K^-}+f^{+}_{s\rightarrow d\bar K^0}\right] 
\label{sprob}
\end{eqnarray}
can be read off the s quark line in Eq.~\ref{fluc} and total meson 
emission from the d quark is equal to that from the u quark, $P_d=P_u$.
\par

The proton's probability composition law~\cite{EHQ,WSK,SMW,WB}
\begin{eqnarray}\nonumber
(1-P_u)(u^0_{\uparrow}\hat u_{\uparrow} +u^0_{\downarrow}\hat u_{\downarrow} 
 +d^0_{\uparrow} \hat d_{\uparrow} +d^0_{\downarrow} \hat d_{\downarrow})
\end{eqnarray}
\begin{eqnarray}
 + u^{0}_{\uparrow}\sum_{q,m} p_m P^{+}_{u_{\uparrow}\rightarrow 
q_{\downarrow} m} 
+u^0_{\downarrow}\sum_{q,m} p_m P^{+}_{u_{\downarrow}\rightarrow 
q_{\uparrow} m} 
+d^0_{\uparrow}\sum_{q,m} p_m P^{+}_{d_{\uparrow}\rightarrow q_{\downarrow} m} 
+d^0_{\downarrow}\sum_{q,m} p_m P^{+}_{d_{\downarrow}\rightarrow q_{\uparrow} m}
\label{rqp}
\end{eqnarray}
is a useful tool in conjunction with Eq.~\ref{fluc} for keeping track of 
chiral fluctuations from valence quark distributions $q^0(x)$. The latter will 
be addressed in Sect.V.  
\par
The splitting of quarks into a Goldstone boson and a recoil quark corresponds 
to a factorization of DIS structure functions that is equivalent to a 
convolution of quark distributions with splitting functions at a scale to be 
discussed in Sect.III.    
From Eqs.~\ref{fluc},\ref{rqp} the unpolarized quark distributions off 
valence quark distributions, $q^0(x)$, and up to second order chiral 
fluctuations, are given by convolution integrals of the general form  
$$P^+\otimes q^0=\int^{1}_{x}{dx_1\over x_1}P^{+}_{q\rightarrow q'GB}
({x\over x_1})q^{0}(x_1),$$\ 
$$q\otimes P^+\otimes q^0
=\int^{1}_{x}{dx_1\over x_1}q_{GB}({x\over x_1})\int^{1}_{x_1}{dx_2\over x_2}
P^{+}_{q\rightarrow q'GB}({x_1\over x_2})q^{0}(x_2),$$
so that 
\begin{eqnarray}\nonumber 
u(x)=Zu^0+P^{+}_{d\rightarrow u\pi ^-}\otimes d^0
+\left({1\over 2}P^{+}_{u\rightarrow u\pi ^0}
+{1\over 6}P^{+}_{u\rightarrow u\eta }\right)\otimes u^0
+u^{0}_{\pi ^+}\otimes P^{+}_{u\rightarrow d\pi ^+}\otimes u^0
\end{eqnarray}
\begin{eqnarray}\nonumber 
+u^{0}_{K^+}\otimes P^{+}_{u\rightarrow s K^+}\otimes u^0
+{1\over 4}u^{0}_{\pi ^0}\otimes P^{+}_{q\rightarrow q\pi ^0}\otimes (u^0+d^0)
+{1\over 36}u^{0}_{\eta }\otimes P^{+}_{q\rightarrow q\eta }\otimes (u^0+d^0)
\end{eqnarray}
\begin{eqnarray}
+{1\over 6}u^{0}_{\pi ^0,\eta }\otimes P^{+}_{q\rightarrow q\pi ^0,\eta  }
\otimes (u^0-d^0),
\label{u}
\end{eqnarray}
\begin{eqnarray}\nonumber
d(x)=Zd^0+P^{+}_{u\rightarrow d\pi ^+}\otimes u^0
+\left({1\over 2}P^{+}_{d\rightarrow d\pi ^0}
+{1\over 6}P^{+}_{d\rightarrow d\eta }\right)\otimes d^0
+d^{0}_{\pi ^-}\otimes P^{+}_{d\rightarrow u\pi ^-}\otimes d^0
\end{eqnarray}
\begin{eqnarray}\nonumber 
+d^{0}_{K^0}\otimes P^{+}_{d\rightarrow s K^0}\otimes d^0
+{1\over 4}d^{0}_{\pi ^0}\otimes P^{+}_{q\rightarrow q\pi ^0}\otimes (u^0+d^0)
+{1\over 36}d^{0}_{\eta }\otimes P^{+}_{q\rightarrow q\eta }\otimes (u^0+d^0)
\end{eqnarray}
\begin{eqnarray}
-{1\over 6}u^{0}_{\pi ^0,\eta }\otimes P^{+}_{q\rightarrow q\pi ^0,\eta  }
\otimes (u^0-d^0),
\label{d}
\end{eqnarray}   
\begin{eqnarray}
s=P^{+}_{u\rightarrow s K^+}\otimes u^0+P^{+}_{d\rightarrow s K^0}\otimes d^0 
+{4\over 9}s^{0}_{\eta }\otimes P^{+}_{q\rightarrow q\eta }\otimes (u^0+d^0). 
\label{s}
\end{eqnarray}
Here the $q^{0}_{GB}$ are the valence quark distributions of the respective 
Goldstone boson.  
\par
Moreover, the $u$ (or $d$) amplitudes from the $\pi ^0$ and $\eta $ 
meson exchanges add coherently to the total antiquark probability amplitude of 
the proton. We know from ref.\cite{EHQ,WSK,SMW} that these interference 
amplitudes, i.e. the last terms in Eq.~\ref{u},\ref{d}, cannot be neglected. 
Here we have approximated the $(\pi ^0,\eta )$ interference terms by using 
equal $\pi ^0$ and $\eta $ splitting functions and quark distributions 
$q_{\pi ^0}\approx q_{\eta }\approx q$ in these mesons so that  
\begin{eqnarray}
\int^{1}_{x}{dx_1\over x_1}[q_{\pi ^0}(x_1)q_{\eta }(x_1)]^{1/2}
\int^{1}_{x_1}[P^{+}_{q\rightarrow q(\pi ^0)}({x_2\over x_1})
P^{+}_{q\rightarrow q(\eta )}({x_2\over x_1})]^{1/2}q^0({x\over x_2})
{dx_2\over x_2}\\  
\sim \int^{1}_{x}{dx_1\over x_1}q(x_1)
\int^{1}_{x_1}P^{+}_{q\rightarrow q(\pi ^0,\eta )}({x_2\over x_1})
q^0({x\over x_2}){dx_2\over x_2}
\label{antis}
\end{eqnarray}
takes the form of a conventional parton distribution. 

\section{Antiquark and spin distributions} 
In chiral dynamics, antiquarks originate only from the Goldstone bosons via 
their quark-antiquark composition, viz. 
\begin{eqnarray}
|\pi ^0\rangle={1\over \sqrt{2}}(\bar u u-\bar d d),\ 
|\eta \rangle={1\over \sqrt{6}}(\bar u u+\bar d d-2\bar s s),\ 
|K^+\rangle=u\bar s, 
\label{qbarq}
\end{eqnarray}
etc. Therefore, {\bf antiquarks are unpolarized}. Small antiquark polarizations 
are consistent with the most recent SMC data~\cite{SMC} so that we expect only 
small corrections if we use $\bar u_\uparrow = \bar u_\downarrow$ in the spin 
fractions $\Delta u = u_\uparrow -u_\downarrow +\bar u_\uparrow 
-\bar u_\downarrow$, etc., i.e. $\Delta s=\Delta s_{sea}$, 
$\Delta \bar u=\Delta \bar d=\Delta \bar s =0$.  
\par
An antiquark in a Goldstone boson which is struck by the virtual photon in DIS 
is linked to a valence quark of the proton by Goldstone boson exchange. 
Therefore, antiquarks appear only in second order of chiral fluctuations.  
Thus, from Eqs.~\ref{fluc},\ref{rqp} the antiquark fractions 
of the proton are given by double convolution integrals of the general form  
$\bar q\otimes P_{q\rightarrow qGB}\otimes q^0$,    

\begin{eqnarray}\nonumber
\bar u(x)=\bar u^{0}_{\pi ^-}\otimes P^{+}_{d\rightarrow u\pi ^-}\otimes d^0
+{1\over 4}\bar u^{0}_{\pi ^0}\otimes P^{+}_{q\rightarrow q\pi ^0}\otimes 
(u^0+d^0)+{1\over 36}\bar u^{0}_{\eta }\otimes P^{+}_{q\rightarrow q\eta }
\otimes (u^0+d^0)
\end{eqnarray}
\begin{eqnarray}
+{1\over 6}\bar u^{0}_{\pi ^0,\eta }\otimes P^{+}_{q\rightarrow q\pi ^0,\eta }
\otimes (u^0-d^0),
\label{ub}
\end{eqnarray}
\begin{eqnarray}\nonumber
\bar d(x)=\bar d^{0}_{\pi ^+}\otimes P^{+}_{u\rightarrow d\pi ^+}\otimes u^0
+{1\over 4}\bar d^{0}_{\pi ^0}\otimes P^{+}_{q\rightarrow q\pi ^0}\otimes 
(u^0+d^0)+{1\over 36}\bar d^{0}_{\eta }\otimes P^{+}_{q\rightarrow q\eta }
\otimes (u^0+d^0)
\end{eqnarray}
\begin{eqnarray}
-{1\over 6}\bar d^{0}_{\pi ^0,\eta }\otimes P^{+}_{q\rightarrow q\pi ^0,\eta }
\otimes (u^0-d^0),
\label{db}
\end{eqnarray}
\begin{eqnarray}
\bar s=\bar s^{0}_{K^0}\otimes P^{+}_{d\rightarrow s K^0}\otimes d^0
+\bar s^{0}_{K^+}\otimes P^{+}_{u\rightarrow s K^+}\otimes u^0
+{4\over 9}\bar s^{0}_{\eta }\otimes P^{+}_{q\rightarrow q\eta }\otimes 
(u^0+d^0). 
\label{sb}
\end{eqnarray}

Again, we have approximated the $(\pi ^0, \eta )$ interference terms, the last 
expression in Eqs.~\ref{ub},\ref{db}, by using equal $\pi ^0$ and $\eta $ 
splitting functions and the same antiquark distributions in these mesons.  

\par
The $\bar s(x)$ and $s(x)$ distributions of Eqs.~\ref{s}, \ref{sb}, 
respectively, are no longer equal unless the s valence antiquark distributions 
in the K mesons, $\bar s^{0}_{K^0,K^+}\sim \delta(1-x)$, are approximated by 
the elastic limit. When spectator quarks are included in addition, but second 
order chiral fluctuations are neglected, then one recovers from Eq.~\ref{s} 
the strange quark distribution, $s(x)$, of some hadronic 
models~\cite{Meln,BMa}.  
\par
Upon generalizing the formalism of~\cite{WSK} also to the polarized  
quark distributions, the probabilities displayed in Eq.~\ref{fluc} and 
Eq.~\ref{rqp} yield  
\begin{eqnarray}
q_{\uparrow}(x)=(1-P_{q})q^{0}_{\uparrow}(x)+\sum_{m,q'} p_m 
P^{+}_{q'\rightarrow q m}\otimes q'^{0}_{\downarrow}+...,
\label{oldq} 
\end{eqnarray}
where the constants $p_m$ are defined below Eq.~\ref{fock} and $m$ denotes a 
Goldstone boson. The corresponding result holds for the other quark helicity. 
The ellipses in 
Eq.~\ref{oldq} denote double convolution terms with $q^{0}_{\uparrow}$ from a
Goldstone boson $m$ that cancel in $q_{\uparrow}-q_{\downarrow}$. The opposite 
quark helicity on the rhs of Eq.~\ref{oldq} means that helicity non-flip 
chiral transitions are neglected and implies the negative sign of all  
chiral contributions to the spin distributions
\begin{eqnarray}
\Delta q(x) = (1-P_{q})\Delta q^0(x) -\sum_{m,q'} p_m P^{+}_{q'\rightarrow q m}
\otimes \Delta q'^0. 
\label{delp}
\end{eqnarray}
The general result involves the polarized splitting functions  
\begin{equation}
P^{-}_{q\rightarrow q' GB}(y)=\int d^2{\bf k}_\perp f^{-}_{q\rightarrow q' GB}
(y,{\bf k}_\perp) 
\label{polspl}
\end{equation}    
and has the same form as Eq.~\ref{delp},    
\begin{eqnarray}
\Delta u(x) = (1-P_u)\Delta u^0(x)
+\int^{1}_{x}{dy\over y}\left[\Delta u^0(y)\left({1\over 2}
P^{-}_{u\rightarrow u\pi ^0}+{1\over 6}P^{-}_{u\rightarrow u\eta }
\right)+\Delta d^0(y)P^{-}_{d\rightarrow u\pi ^-}\right],
\label{delu}
\end{eqnarray}
\begin{eqnarray}
\Delta d(x) = (1-P_d)\Delta d^0(x)
+\int^{1}_{x}{dy\over y}\left[\Delta d^0(y)\left({1\over 2}
P^{-}_{d\rightarrow d\pi ^0}+{1\over 6}P^{-}_{d\rightarrow d\eta }
\right)+\Delta u^0(y)P^{-}_{u\rightarrow d\pi ^+}\right],
\label{deld}
\end{eqnarray}
\begin{equation}
\Delta s(x)=\int^{1}_{x}{dy\over y}\left[\Delta d^0(y)
P^{-}_{d\rightarrow sK^0}({x\over y})
+\Delta u^0(y)P^{-}_{u\rightarrow s K^+}({x\over y})\right], 
\label{dels}
\end{equation}
except for the replacement of $-P^+$ by the corresponding polarized splitting 
function $P^-$. 
\par
Due to the minus sign in the ${\bf k}^2_{\perp}$ term of the polarized 
splitting functions in Eq.~\ref{splf} which corresponds to the subtracted 
terms in Eq.\ref{delp}, the general reduction of 
$\Delta q^{0}$ by chiral fluctuations in lowest order is the crucial property 
responsible for the success of chiral field theory for the proton spin 
fractions. A comparison of Eq.~\ref{delp} with Eqs.\ref{delu},\ref{deld},
\ref{dels}, shows that only the unpolarized splitting function 
occurs in the polarized and unpolarized spin-flavor fractions of 
refs.\cite{EHQ,WSK,SMW,CLi,WB}, and the negative sign of $P^{-}\approx -P^{+}$ 
is critical for the reductions of the spin fractions by the chiral fluctuations 
and the remarkable success of these models. Thus, when these 
$\Delta q(x)$ are integrated over Bjorken x, the lowest moments 
(which factorize into products of moments as we discuss in the next section) 
reproduce precisely the structure of the results for the spin 
fractions~\cite{WSK,WB} 
\begin{eqnarray}\nonumber
\Delta u=(1-P)\Delta u^0-\left({2\over 3}\Delta u^0+\Delta d^0\right)a
\left(1+{\epsilon\over \sqrt{3}}\right)^2,\\\nonumber
\Delta d=(1-P)\Delta d^0-\left({2\over 3}\Delta d^0+\Delta u^0\right)a
\left(1+{\epsilon\over \sqrt{3}}\right)^2,\\
\Delta s=-({2\over 3}\Delta d^0+\Delta u^0)a\left(1-{2\epsilon\over 
\sqrt{3}}\right)^2.
\label{delo}
\end{eqnarray}
The pure spin flip approximation is valid at high quark momentum or, upon 
comparing $P^{-}=-P^{+}$ of Eq.~\ref{delp} with the numerator of the splitting 
functions in Eq.~\ref{splf}, when the quark masses are negligible compared to 
the transverse quark momentum $<|{\bf k}|_{\perp}>$, i.e. current quarks are 
the relevant degrees of freedom. The factor 
$2/3$ in Eq.~\ref{delo} comes from adding ${1\over 2}+{1\over 6}$ in 
Eqs.~\ref{delu},\ref{deld} using the 
same splitting function for the $\pi ^0$ and $\eta $ mesons because the SU(3) 
breaking along the $\lambda _8$ direction (which is the same as for the standard
baryon and meson mass splittings) does not suppress the $\eta $ contribution. 
In particular, taking the elastic limit for the quark distributions in the 
mesons, $q(x_2)\sim \delta(1-x_2)$, as well as the elastic-limit delta 
function for the valence quark distributions $q^0(x_1)$, but keeping all 
second order chiral fluctuations, 
yields all our previous integrated spin-flavor fractions~\cite{WSK,WB} and 
Eq.~\ref{delo} in particular. Comparing the unpolarized quark and antiquark 
fractions already yields the identification 
\begin{eqnarray}
a(1+{\epsilon \over \sqrt{3}})^2=\int^{1}_{0}dxP^{+}_{u\rightarrow d\pi ^+},\  
a(1-{2\epsilon\over \sqrt{3}})^2=\int^{1}_{0}dxP^{+}_{u\rightarrow s K^+}, 
\label{old}
\end{eqnarray}
where $a$ is the chiral coupling strength and $\epsilon $ the SU(3) breaking 
parameter of~\cite{WSK,WB}. For other chiral models such as~\cite{SMW,CLi} 
similar identifications can be made.  
\par
This detailed comparison indicates that the success of these chiral models can 
be attributed to parametrizations that rely on pure helicity flip chiral 
transitions. Since the helicity non-flip chiral transition probabilities 
depend on the quark masses, these fluctuations are negligible only when the 
quark masses are small compared to the relevant transverse momentum range. 
This is not the case for constituent quarks. Therefore, chiral quark models 
for the spin fractions have to include helicity non-flip fluctuations. When  
this is done as in ref.~\cite{SW} then the spin fractions disagree with the 
data.  

\section{Moments}

The proton spin observables $\Delta q$ are the lowest moments of, or integrals 
over, the polarized structure functions. Also, the constraint imposed on 
the quark distributions by the momentum involves the first moment 
$\int^{1}_{0}xq(x)dx$ of the quark distributions. Therefore, in the context of 
DIS and chiral dynamics, it is mandatory to consider moments of convolutions 
of quarks or antiquarks with splitting functions.     

It is straightforward to verify that the 
integral of a convolution, e. g. 
\begin{eqnarray}
\int^{1}_{0}\bar q\otimes P \otimes q f(x)dx=\int^{1}_{0}\bar q(x)f(x)dx
\cdot\int^{1}_{0}P(x)f(x)dx\cdot\int^{1}_{0}q(x)f(x)dx,
\label{momfac}
\end{eqnarray}
factorizes provided $f$ is a multiplicative function, i.e. $f(xy)=f(x)f(y)$, 
such as the power law $x^n$ that defines the n'th moment. In other words,  
the moment of a convolution is the product of the moments of the components of 
the convolution. 

Applying this general observation to the quark, antiquark and spin 
distributions of Eqs.~\ref{u},\ref{s},\ref{ub},\ref{sb},\ref{delu},
\ref{deld},\ref{dels} yields for the 
moments $q_n\equiv\int^{1}_{0}x^nq(x)dx$ the following relations 
\begin{eqnarray}\nonumber
u_n=a_n u_{n}^{0}+b_n d_{n}^{0},\qquad d_n=a_n d_{n}^{0}+b_n u_{n}^{0},
\end{eqnarray}
\begin{eqnarray}
s_n=P^{+,n}_{u\rightarrow sK^+}u_{n}^{0}+P^{+,n}_{d\rightarrow sK^0}d_{n}^{0} 
+{4\over 9}s_{\eta ,n}^{0}P^{+,n}_{q\rightarrow q\eta }(u_{n}^{0}+d_{n}^{0}),  
\label{qmom}
\end{eqnarray}
\begin{eqnarray}\nonumber
\bar{u}_n=\bar{u}^{0}_{\pi -,n}P^{+,n}_{d\rightarrow u\pi ^-}d^{0}_{n}
+{1\over 4}\bar{u}^{0}_{\pi ^0,n}P^{+,n}_{\pi ^0}(u^{0}_{n}+d^{0}_{n})
+{1\over 36}\bar{u}^{0}_{\eta ,n}P^{+,n}_{\eta }(u^{0}_{n}+d^{0}_{n})
+{1\over 6}\bar{u}^{0}_{n}P^{+,n}(u^{0}_{n}-d^{0}_{n}),
\end{eqnarray}
\begin{eqnarray}
\bar{s}_n=\bar{s}^{0}_{K^+,n}P^{+,n}_{u\rightarrow sK^+}u_{n}^{0}
+\bar{s}^{0}_{K^0,n}P^{+,n}_{d\rightarrow sK^0}d_{n}^{0} 
+{4\over 9}\bar{s}_{\eta ,n}^{0}P^{+,n}_{q\rightarrow q\eta }
(u_{n}^{0}+d_{n}^{0}),   
\label{qsm}
\end{eqnarray}
where the $P^{+,n}$ are the moments of the unpolarized splitting functions and 
the $q^{0}_{n}, \bar{s}^{0}_{n}$ are those of the proton and K meson  
valence quark distributions. The coefficients $a_n, b_n$ are given by 
\begin{eqnarray}\nonumber
a_n=Z+{1\over 2}P^{+,n}_{\pi ^0}+{1\over 6}P^{+,n}_{\eta }
+u^{0}_{\pi ^+,n}P^{+,n}_{u\rightarrow d\pi ^+}\\
+{1\over 4}u^{0}_{\pi ^0,n}P^{+,n}_{\pi ^0}
+{1\over 36}u^{0}_{\eta ,n}P^{+,n}_{\eta }+{1\over 6}u^{0}_{n}P^{+,n},
\end{eqnarray} 
\begin{eqnarray}
b_n=P^{+,n}_{d\rightarrow u\pi ^-}
+{1\over 4}u^{0}_{\pi ^0,n}P^{+,n}_{\pi ^0}
+{1\over 36}u^{0}_{\eta ,n}P^{+,n}_{\eta }-{1\over 6}u^{0}_{n}P^{+,n}. 
\label{coeff}
\end{eqnarray} 

Similarly, from the $\Delta q$ in Eq.~\ref{dels} the spin dependent moments 
are obtained as  
\begin{eqnarray}
\Delta u_n=(1-P_u)\Delta u^{0}_{n}+\left[{1\over 2}P^{-,n}_{\pi ^0}
+{1\over 6}P^{-,n}_{\eta }\right]\Delta u^{0}_{n}
+P^{-,n}_{d\rightarrow u\pi ^-}\Delta d^{0}_{n}, 
\end{eqnarray}
\begin{eqnarray}
\Delta d_n=(1-P_d)\Delta d^{0}_{n}+\left[{1\over 2}P^{-,n}_{\pi ^0}
+{1\over 6}P^{-,n}_{\eta }\right]\Delta d^{0}_{n}
+P^{-,n}_{u\rightarrow d\pi ^+}\Delta u^{0}_{n}, 
\end{eqnarray}
\begin{eqnarray}
~\Delta s_n=P^{-,n}_{u\rightarrow sK^+}\Delta u^{0}_{n}
+P^{-,n}_{d\rightarrow sK^0}\Delta d^{0}_{n} 
\end{eqnarray}       
to be compared with Eq.~\ref{delo} for $n=0$.  
\par 
As a consequence of the factorization of moments of convolutions, the equality 
of the integrated $\bar{s}=s$ of the proton follows immediately from 
Eqs.~\ref{qmom},\ref{qsm} using $\bar{s}^{0}_{0}=1$ for the valence s antiquark 
fraction in the K mesons. In order to avoid more parameters and complications 
we shall use the elastic delta function approximation for the valence quarks 
and antiquarks in the Goldstone bosons in the numerical evaluations.  
\par

\section{QCD Constraints and Results}

There are numerous attempts to describe DIS structure functions based on quark 
models. These quark models provide initial quark distributions at the scale 
$\Lambda_{QCD}$ as a rule. For the NQM~\cite{Mo} and a light-cone quark model 
(LCQM~\cite{W}) the up quark distributions are exhibited in Fig.1. 
The subsequent $Q^2$-evolution of structure 
functions follows from perturbative QCD, but relies on such initial shapes 
which contain the non-perturbative aspects of the bound state. The low-energy 
scale of a quark model is often treated as an adjustable parameter which 
is usually too small for a perturbative evolution to be valid. It is fair to 
say that, even with chiral fluctuations, quark models fall short of explaining 
the deep inelastic structure functions and the spin of the proton~\cite{SW}, 
quite apart from rather significant uncertainties in their evolution from 
$\Lambda _{QCD}$ to about $\Lambda _{\chi }$.  

The success of chiral fraction models based on pure helicity flip chiral 
fluctuations suggests that we construct initial quark distributions for the 
case where current quarks are the relevant degrees of freedom. Upon comparing  
with the momentum dependence of the dynamical quark mass discussed in the 
Introduction, the relevant momentum scale is about ${2\over 3}$ GeV to 
$\Lambda _{\chi }$ rather than $\Lambda _{QCD}$ of the NQM. 
It is important to realize in this context that the use of the spin fractions 
$\Delta q^0$ from the spin-flavor proton wave function of the NQM in 
Eq.~\ref{delo} is not only justified at scale $\Lambda _{QCD}$ but at 
$\Lambda _{\chi }$ as well because the (for $\Delta s^0=0$) equivalent 
isotriplet and octet axial charges $g^{(3)}_{A}, g^{(8)}_{A}$ are conserved in 
the chiral limit and scale independent.   

Instead of the NQM we use the BBS approach to construct valence quark 
distributions. In contrast to BBS, we impose the quark 
counting rules directly on the valence quark distributions. There are no 
parametrizations for strange quarks or antiquarks which we expect to be 
generated by chiral fluctuations. It is an immediate consequence that initial 
parton distributions where the strange quarks contribute nothing (for some 
examples, see~\cite{GR}) are ruled out by chiral field theory. 
\par
The constraints
\begin{eqnarray} 
u^{0}_{0}=\int^{1}_{0}u^0(x)dx=2,\quad  d^{0}_{0}=\int^{1}_{0}d^0(x)dx=1\quad 
\label{num}
\end{eqnarray}
on valence quark numbers are 
characteristic features of quark models, such as the nonrelativistic NQM and 
light cone quark models. Moreover, the quark distributions satisfy  
$q^{0}\rightarrow 0$ as $x\rightarrow 0$ in these models.    
Hence for illustration purposes only, we parametrize such valence 
quark distributions $q^{0}(x)$ in terms of the shapes 
$x^{\alpha }(1-x)^{\beta }$ with a common {\bf positive} exponent $\alpha $ and 
$\beta =$3, or 5.   
As a consequence of the positive exponent, fewer non-leading terms are needed, 
which saves more than half of BBS's parameters, viz. 
\begin{eqnarray}\nonumber 
u^{0}_{\uparrow}(x) = Ax^{\alpha }(1-x)^3,\quad
u^{0}_{\downarrow}(x) = Cx^{\alpha }(1-x)^5,\quad
\end{eqnarray}
\begin{eqnarray}
d^{0}_{\uparrow}(x) = {A\over 5}x^{\alpha }(1-x)^3,\quad
d^{0}_{\downarrow}(x) = C'x^{\alpha }(1-x)^5,  
\label{qmv}
\end{eqnarray}   
where the anti-parallel helicity quark distributions are suppressed by the  
factor $(1-x)^2$ and the exponent $\beta =$3 for $u^{0}_{\uparrow}$ and 
$d^{0}_{\uparrow}$ reflects the quark counting rule for the proton that 
guarantees the correct falloff of the leading form factors with $q^2$. 
The choice $A/5$ for the leading coefficient of 
$d^{0}_{\uparrow}$ in Eq.~\ref{qmv} is motivated by the SU(6) structure of 
the NQM and its valence quark distributions for large $x$. When the other 
ratios of coefficients are also chosen as in the SU(6) spin-flavor wave 
function of the proton (cp. Eq.13 of ~\cite{WSK}),
\begin{eqnarray}
C={A\over 5},\ C'={2\over 5}A,\  
\end{eqnarray}
then we obtain 
\begin{eqnarray}
A=g^{(3)}_{A}\Pi^{6}_{j=1}(j+\alpha )/6^3,\qquad 
\alpha=\left[{161\over 4}+{36\Delta d^0\over g^{(3)}_{A}}\right]^{1/2}
-{11\over 2}\approx {1\over 2}
\end{eqnarray}
for an initial value $\Delta d^0\sim -0.2$ which increases to -0.4 by chiral 
fluctuations. The corresponding up quark distribution with chiral 
fluctuations is close to the dot-dashed line in Fig.1, where the exponent 
$\alpha =1/2$ and $\Delta u^0=4/3$ are chosen to simulate the NQM.       
It is straightforward to express the coefficients C, C' in Eq.~\ref{qmv} for 
that case in terms of A and $\alpha $ from the two constraints of 
Eq.~\ref{num}. The formulas for the coefficients in Eq.~\ref{qmv} are given in 
the Appendix and their numerical values are displayed in Table 2. Upon 
imposing the observed values for 
$g_{A}^{(3)},\ g_{A}^{(8)}$, or equivalently $\Delta u^0, \Delta d^0$ as well, 
the relation $\Delta u^0-5\Delta d^0=3$ is obtained, 
which is satisfied by the NQM values $\Delta u^0=4/3, \Delta d^0=-1/3$. Thus,  
for fixed values of the exponent $\alpha $ and $\Delta u^0$, which are not 
adjusted to any observable, the chiral strength is adjusted to fit 
$g_{A}^{(3)}$ with chiral fluctuations, yielding the results in Figs.1,2. The 
case with a smaller 
$\Delta u^0=1.1$ is supposed to simulate a relativistic quark model. The price 
one pays for selecting a positive $x$-power is that the resulting 
quark distributions shown as dot-dashed line in Fig.1 for exponent 
$\alpha =1/2$ and solid lines in Fig.2 become unrealistically small for  
$x\rightarrow 0$, which is a typical quark model feature. Clearly, these quark 
distributions have a lower peak than those of the NQM and LCQM, and it is 
shifted to lower $x$ values, features that make them look like evolved from 
quark models in Fig.1 at $\Lambda _{QCD}$ to a higher scale. The comparison of 
the dotted curve in Fig.2 which is the associated valence $u^{0}(x)$ quark 
distribution without chiral fluctuations with the upper solid line shows the 
typical size of chiral fluctuations; they shift the peak to lower $x$ but 
increase its magnitude for the u quark, while it is lowered for the d quark.  
\par
The integrated spin observables obtained with chiral fluctuations 
are given in Table 1 for the cases shown in Figs.1,2,3. Note that for the 
quark model simulations with exponent $\alpha =$1/2, despite fitting only one 
parameter, the chiral strength $\alpha _{\chi }$, to the axial charge of the 
nucleon, most of the other spin observables are not too far off. In Fig.3, 
though, the exponent $\alpha =$0.1 is deliberately chosen small to get closer 
to the BBS result for $u(x)$. For the smaller exponent $\alpha =0.1$, the 
total quark momentum constraint $u^{0}_{1}+d^{0}_{1}=0.52$ is imposed instead 
of the quark number constraints of Eq.~\ref{num}, which also helps getting 
(the dotted line in Fig.3) closer to the BBS (dot-dashed) result, except at 
small $x$. For this case, nearly all integrated spin fractions in Table 1 are 
already in good 
agreement with the data, a success for which chiral fluctuations are essential 
in conjunction with current quark masses. Clearly, all these quark 
distributions become unrealistically small for $x\leq$ 0.1. 
\par
When the light up 
and down quarks are taken to be constituent with $m_q=m_N/3=313$ MeV, and the 
strange quark with $m_s=m_{\Sigma }-m_N+m_q=567$ MeV, then most numerical 
results become worse; the chiral strength has to be 
increased, while the proton spin $\Delta \Sigma $ moves up to about 0.5, 
confirming the results of ref.~\cite{SW} 
\par
Since the recent HERA data~\cite{HERA} clearly show the rise of the unpolarized 
$F^{p}_{2}(x)$ at small $x$, which the quark model-type distributions in 
Figs.1,2,3 fail to reproduce, in the next step we incorporate Regge behavior 
imposing a negative common (isoscalar) power $x^{-\alpha _R}$ on the valence 
distributions. Just like BBS, we find that more non-leading terms must be 
introduced to make the proton spin fractions $\Delta q$ and the Gottfried sum 
rule (GSR) convergent.  
The minimal Regge-type parametrization is given by 
\begin{eqnarray}\nonumber 
~u^{0}_{\uparrow}(x) = x^{-\alpha _R}[A+B(1-x)](1-x)^3,\quad
u^{0}_{\downarrow}(x) = x^{-\alpha _R}[C+D(1-x)](1-x)^5,\qquad 
\end{eqnarray}
\begin{eqnarray}
~d^{0}_{\uparrow}(x) = x^{-\alpha _R}\left[{A\over 5}+B'(1-x)\right](1-x)^3,
\quad
d^{0}_{\downarrow}(x) = x^{-\alpha _R}[C'+D'(1-x)](1-x)^5. 
\label{rv}
\end{eqnarray}   
For $\Delta u^0(x), \Delta d^0(x)$ to be finite for $x\rightarrow 0$, the 
constraints 
\begin{equation}
A + B = C + D,\quad {A\over 5} + B' = C' + D'\quad
\end{equation}
are required. 
For the Gottfried sum rule to be convergent, or isospin symmetry for 
$x\rightarrow 0$, requires 
\begin{equation}
A + B + C + D = {A\over 5} + B' + C' + D'.\quad 
\end{equation}
Moreover, we impose fixed values for $\Delta u^0, \Delta d^0$ and the GSR, or 
equivalently $u^{0}_{0}-d^{0}_{0}$, and the total quark momentum 
$u^{0}_{1}+d^{0}_{1}$. Again, it is straightforward but tedious to determine 
the coefficients in Eq.~\ref{rv} from these conditions; they are given 
analytically in the Appendix and numerically in Table 2.  
\par
Chiral fluctuations will not change the Regge exponent in the dressed quark 
distributions because of the asymptotic behavior of convolutions, e.g. from   
$q^{0}(x) \sim x^{-\alpha _R}$ we obtain 
\begin{eqnarray}
P^{\pm}\otimes q^0 \sim x^{-\alpha _R}\int_{0}^{1}dt P^{\pm}t^{-1-\alpha _R},
\quad   x\rightarrow 0, 
\end{eqnarray}
where the moments of the splitting functions remain finite because the 
Gaussian in Eq.~\ref{fzff} goes to zero faster than any power at the 
endpoints.    
\par
We adopt the Regge 
exponent $\alpha _R=1.12$ of BBS. Then we adjust the chiral strength to the 
spin $\Delta \Sigma $ of the proton roughly, vary  
$\Delta u^0-\Delta d^0, \Delta u^0+\Delta d^0$ (which changes $\Delta \Sigma $ 
a bit) to match the observed values for $g_{A}^{3}, g_{A}^{8}$ with chiral 
fluctuations, impose $u^{0}_{0}-d^{0}_{0}$ to match the GSR and 
$u^{0}_{1}+d^{0}_{1}\approx 1/2$ of the total quark momentum. The results are 
displayed as solid lines in Figs.3-8. The d quark 
distribution in Fig.4 deviates more from BBS, but the s quark distribution in  
Fig.5 (solid line) predicted by chiral fluctuations is again close to the BBS  
(dot-dashed) curve, whereas the quark model case with exponent $\alpha =$0.1 
is much too small at small $x$. The spin dependent $\Delta q$ distributions in 
Figs.6-8 show a similar pattern. The polarized proton structure function 
shown in Fig.9 is crudely evolved by a multiplicative 
$1-\alpha _s/\pi\sim 0.92$ to the data at $-q^2=$10 GeV$^2$ as in~\cite{BBS}. 
The ratio of unpolarized 
$F^{n}_{2}/F^{p}_{2}\rightarrow 3/7$ as $x\rightarrow 1$ with and without 
chiral fluctuations. Finally, the antiquark ratio $\bar u/\bar d$ in Fig.10 is 
not even approximately constant.  
           
\section{Conclusions}  
A comparison of the chiral field theory of quark structure functions  
with previous chiral models~\cite{EHQ,WSK,SMW,CLi,WB} for the spin fractions 
$\Delta q$ shows unambiguously that the remarkable success of these 
models is due to parametrizations that correspond to neglecting helicity 
non-flip chiral fluctuations in the (integrated) polarized splitting functions 
which are sizable for constituent quarks. Thus, {\bf standard} chiral 
dynamics can explain the spin observables of 
the proton only for the case where the dynamical quark masses are already 
small. Furthermore, the present analysis sheds light on the failure of 
standard chiral dynamics at the scale $\Lambda _{QCD}$ involving constituent 
quarks~\cite{SW} whose larger masses significantly reduce the negative chiral 
contributions to the spin fractions. When quark distributions  
are constructed according to the quark counting rules of pQCD in conjunction 
with constraints from the SU(6) spin-flavor wave function of the proton, 
the GSR and total quark momentum fraction, then chiral dynamics 
generates sea quark distributions of correct magnitude and shapes and predicts 
reasonable strange quark distributions in particular so that the spin 
fractions of the proton also agree with the data. 

While chiral dynamics with the (ad-hoc) pure spinflip prescription  
appears to be successful in understanding the proton spin fractions, its  
justification in the context of chiral quark models may be provided by 
instanton dynamics of zero-mode quarks. In fact, 't Hooft's~\cite{GH} 
interaction maintains pure left- to right-handed quark transitions, while light 
quarks pick up a dynamical mass upon propagating in the instanton background 
field.~\cite{DKZ} However, axialvector coupling of constituent quarks with 
Goldstone bosons in standard chiral field theory is inconsistent with the 
proton spin fraction data.  
  
\section{Appendix}
The coefficients of the valence quark distributions for the cases of Fig.1 
(dot-dashed line) and Fig.2 (upper solid and dotted lines) with $\alpha =$1/2 
are given by 
\begin{eqnarray}\nonumber
A={2+\Delta u^0\over 12}\Pi_{j=1}^{4}(j+\alpha ),\quad  
C={2-\Delta u^0\over 240}\Pi_{j=1}^{6}(j+\alpha ),
\end{eqnarray}
\begin{eqnarray} 
C'={8-\Delta u^0\over 1200}\Pi_{j=1}^{6}(j+\alpha ), 
\label{coef}
\end{eqnarray}
those for Fig.3 (dotted line) with $\alpha =$0.1 and total quark momentum 
constraint are 
\begin{eqnarray}\nonumber
A={10\over 147}[(u^{0}_{1}+d^{0}_{1})(5+\alpha )
+(\Delta u^0+\Delta d^0)(1+\alpha )]\Pi_{j=2}^{4}(j+\alpha ),
\end{eqnarray}
\begin{eqnarray}\nonumber 
C={A\over 20}(6+\alpha )(5+\alpha ) 
-{\Delta u^0\over 120}\Pi_{j=1}^{6}(j+\alpha ),
\end{eqnarray}
\begin{eqnarray} 
\ C'={A\over 100}(6+\alpha )(5+\alpha )
-{\Delta d^0\over 120}\Pi_{j=1}^{6}(j+\alpha ),
\end{eqnarray}
while those of the Regge (solid lines in Figs.3-8) and BBS (dot-dashed lines 
in Figs.2-8) are given by  
\begin{eqnarray}\nonumber
A={5\over 48}\left(\Delta u^0-\Delta d^0+u^{0}_{0}-d^{0}_{0}\right)
\Pi_{j=2}^{5}(j-\alpha _R),\quad
\end{eqnarray} 
\begin{eqnarray}\nonumber
{-40B\over A}[(12-\alpha _R)(2-\alpha _R)+(7-\alpha _R)(8-\alpha _R)
+30]=[6(6-\alpha _R)(7-\alpha _R)(8-\alpha _R)\\\nonumber
+16(7-\alpha _R)(8-\alpha _R)+1200+(2-\alpha _R)(732-118\alpha _R
+6\alpha _{R}^{2})]\\
+{1\over 6}\Pi_{j=3}^{7}(j-\alpha _R)\left[(u^{0}_{1}+d^{0}_{1})
(8-\alpha _R)+(\Delta u^0+\Delta d^0)(2-\alpha _R)\right], 
\end{eqnarray}
\begin{eqnarray}\nonumber 
40C=-{\Delta u^0\over 3}\Pi_{j=2}^{7}(j-\alpha _R)+8B(12-\alpha _R)\\
+{A\over 5}[4(7-\alpha _R)(6-\alpha _R)+732-118\alpha _R+6\alpha _{R}^{2}],\\
\nonumber
\ 40C'=-{\Delta d^0\over 3}\Pi_{j=2}^{7}(j-\alpha _R)+8B(12-\alpha _R)\\
+{A\over 5}[732-118\alpha _R+6\alpha _{R}^{2}-4(7-\alpha _R)(6-\alpha _R)].
\end{eqnarray}
\begin{eqnarray}
B'={4\over 5}A + B,\qquad D' = A + B - C'.
\end{eqnarray}
\vfill\eject

{\bf Table 1} Quark Spin Fractions of the Proton for Figs.1-8 

$$
\offinterlineskip \tabskip=0pt 
\vbox{ 
\halign to 1.0\hsize 
   {\strut
  \vrule#                          
   \tabskip=0pt plus 30pt
 & \hfil #  \hfil                  
 & \vrule#                         
 & \hfil #  \hfil                  
 & \vrule#                         
 & \hfil #  \hfil                  
 & \vrule#                         
 & \hfil #  \hfil                  
 & \vrule#                         
 & \hfil #  \hfil                  
 & \vrule#                         
 & \hfil #  \hfil                  
   \tabskip=0pt                    %
 & \vrule#                         
  \cr                             
\noalign{\hrule}
&  && Data E143~\cite{E143}&& $\alpha _{\chi }=$0.39 && $\alpha _{\chi }=$0.3 
&& $\alpha _{\chi }=$0.42 && $\alpha _{\chi }=$0.4 &\cr
& &&at 3 GeV$^2$ && $\alpha ={1\over 2}$ && $\alpha ={1\over 2}$ && $
\alpha =0.1$ && $\alpha _R=1.12$ &\cr
& && Data SMC~\cite{SMC} &&$\Delta u^0={4\over 3}$ &&$\Delta u^0=$1.1&&  &&  
&\cr
& &&at 5 GeV$^2$  &&Figs.1,2 && Figs.2 && Figs.3-8 &&Fig.3-8 &\cr
\noalign{\hrule}
& $\Delta u$ &&0.84$\pm$0.05  && 0.82 && 0.86 && 0.86 && 0.82 &\cr
&            &&0.82$\pm$0.02  &&      &&      &&      &&      &\cr
& $\Delta d$ && -0.43$\pm$0.05 && -0.44 && -0.40 &&-0.40 &&-0.44 &\cr
&            && -0.43$\pm$0.02 &&       &&       &&      &&      &\cr
& $\Delta s$ && -0.08$\pm$0.05 && -0.04 && -0.02 &&-0.07 &&-0.11 &\cr
&            && -0.10$\pm$0.02&&        &&        &&      &&      &\cr    
& $\Delta \Sigma $ && 0.30$\pm$0.06 && 0.344 && 0.44 && 0.39 && 0.27 &\cr
&                  && 0.29$\pm$0.06 &&       &&      &&      &&      &\cr       
& $\Delta_3/\Delta_8$ && 2.09$\pm$0.13 && 2.76 && 2.06 && 2.08 && 2.11 &\cr
& $g^{(3)}_A$&& 1.2601$\pm$0.0025~\cite{PDG}&&1.254&& 1.263&&1.254&&1.254&\cr
& ${\cal F}/{\cal D}$ && 0.575$\pm$0.016 && 0.52 && 0.53 &&0.59 &&0.58 &\cr
& $I_G$      && 0.235$\pm$0.026 && 0.283 && 0.304 && 0.244 &&0.286 &\cr
\noalign{\hrule}
}}$$

\vfill\eject

{\bf Table 2} Coefficients of Valence Quark Distributions in 
Eqs.~\ref{qmv},\ref{rv} for Figs.1-8 

$$
\offinterlineskip \tabskip=0pt 
\vbox{ 
\halign to 1.0\hsize 
   {\strut
  \vrule#                          
   \tabskip=0pt plus 30pt
 & \hfil #  \hfil                  
 & \vrule#                         
 & \hfil #  \hfil                  
 & \vrule#                         
 & \hfil #  \hfil                  
 & \vrule#                         
 & \hfil #  \hfil                  
 & \vrule#                         
 & \hfil #  \hfil                  
 & \vrule#                         
 & \hfil #  \hfil                  
   \tabskip=0pt                    %
 & \vrule#                         
  \cr                             
\noalign{\hrule}
&  && BBS~\cite{BBS} && $\alpha _{\chi }=$0.39 && $\alpha _{\chi }=$0.3 
&& $\alpha _{\chi }=$0.42 && $\alpha _{\chi }=$0.4 &\cr
& && Figs.2-8 && $\alpha ={1\over 2}$ && $\alpha ={1\over 2}$ && $
\alpha =0.1$ && $\alpha _R=1.12$ &\cr
& && &&$\Delta u^0={4\over 3}$ &&$\Delta u^0=$1.1&&  &&  
&\cr
& && &&Figs.1,2 && Figs.2 && Figs.3-8 &&Figs.3-8 &\cr
\noalign{\hrule}
& A && 3.784 && 16.406 && 15.258 && 8.091 && 5.565 &\cr
& B && -3.672 && - && - && - && -5.551 &\cr
& B' && -0.645 && - && - && - && -1.099 &\cr
& C && 2.004 && 5.866 && 7.918 && -0.354 && -1.047 &\cr
& C' && 3.230 && 11.730 && 12.141 && 2.974 && 0.920 &\cr
& D && -1.892 && - && - && - && 1.061 &\cr
& D' && -3.118 && - && - && - && -0.906 &\cr
\noalign{\hrule}
}}$$

\begin{figure}
\vglue 1in
\centerline{\psfig{figure=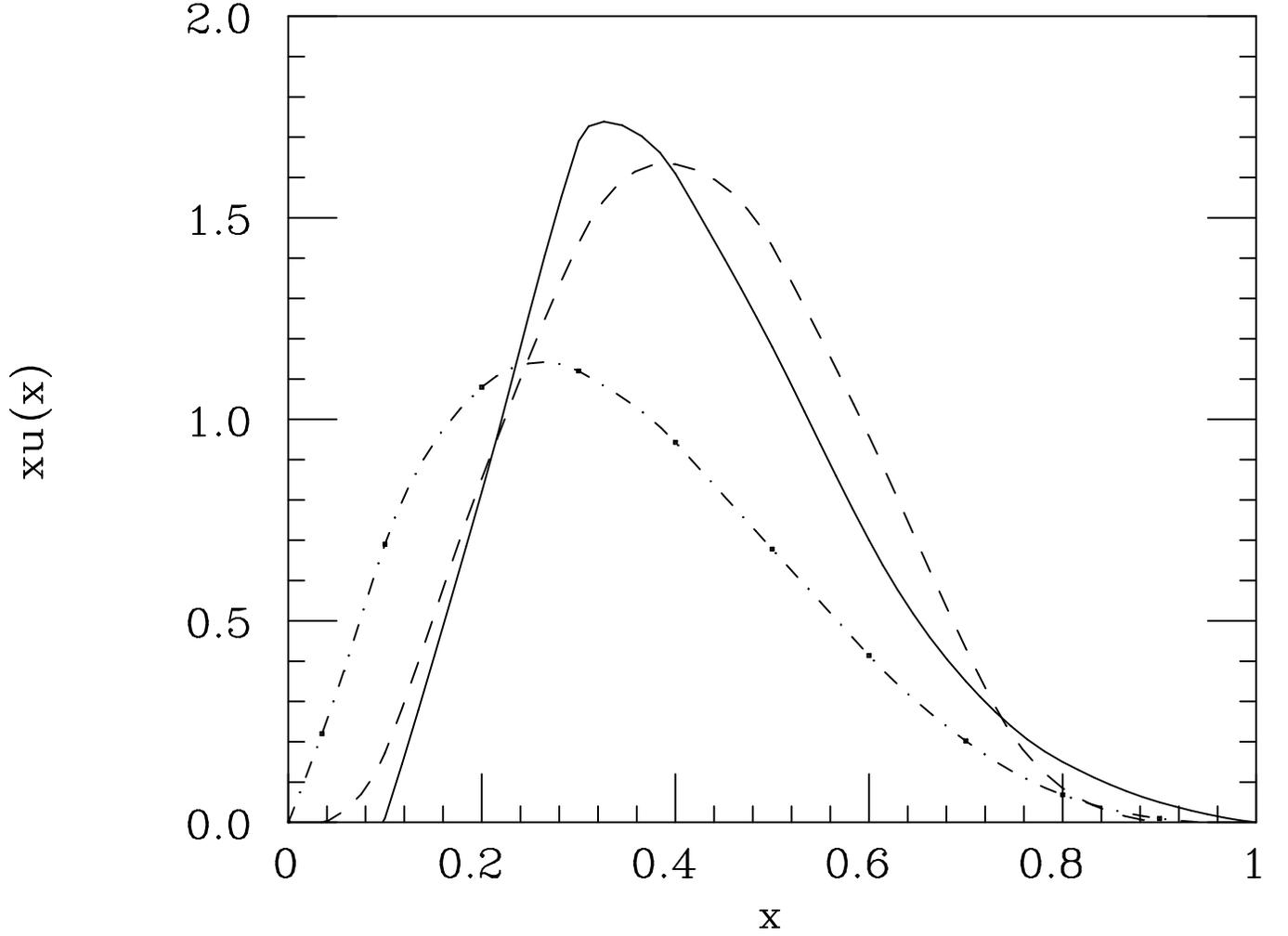,width=6in}}
\vglue 1in
\caption{Quark distribution $xu(x)$ of NQM [18] (solid), LCQM [11] (dashed),
both at $\Lambda _{QCD}$, case of 3rd column of Table 1 (dot-dashed) with 
chiral fluctuations at $\Lambda _{\chi }$.}
\end{figure}

\begin{figure}
\vglue 2in
\centerline{\psfig{figure=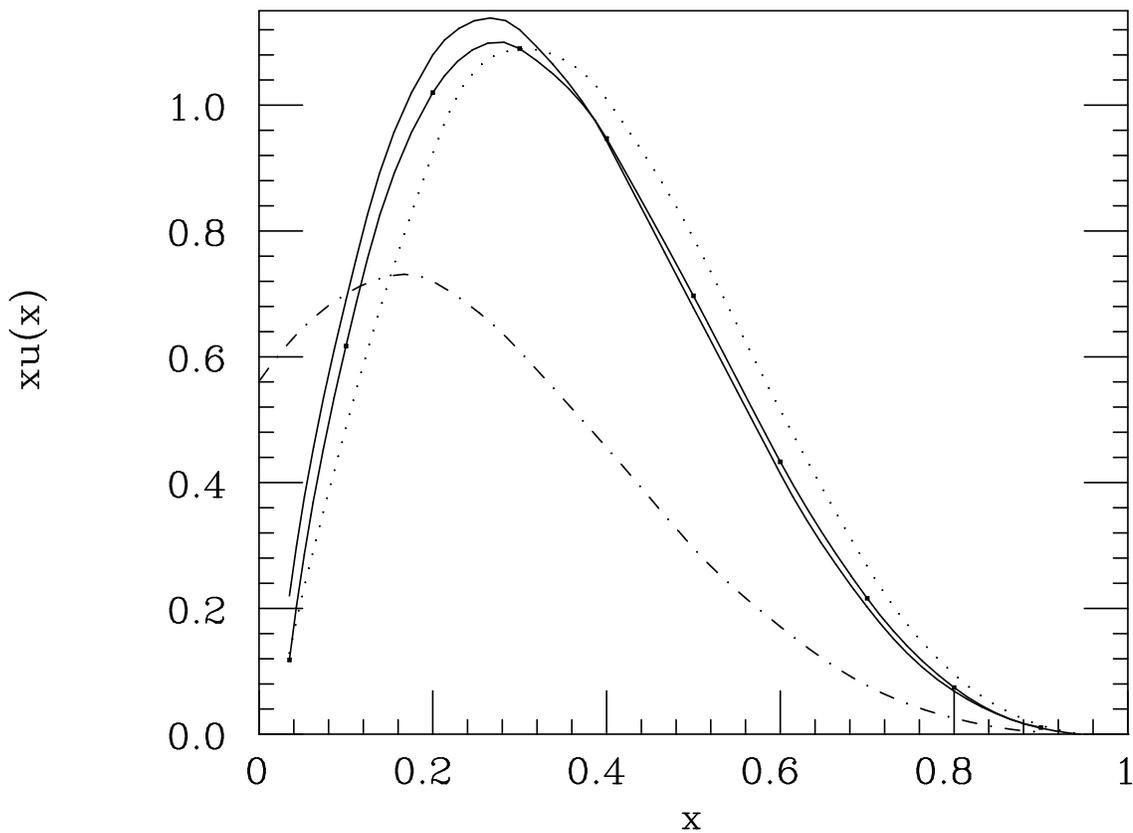,width=5in}}
\caption{Quark distribution $xu(x)$ of 3rd column of Table 1 (top solid) with 
corresponding valence $xu^0(x)$ (dotted) without chiral fluctuations, case of 
4th column of Table 1 (lower solid) compared with BBS [12] (dot-dashed).}
\vglue 1in
\end{figure}

\begin{figure}[h]
\vglue 2in
\centerline{\psfig{figure=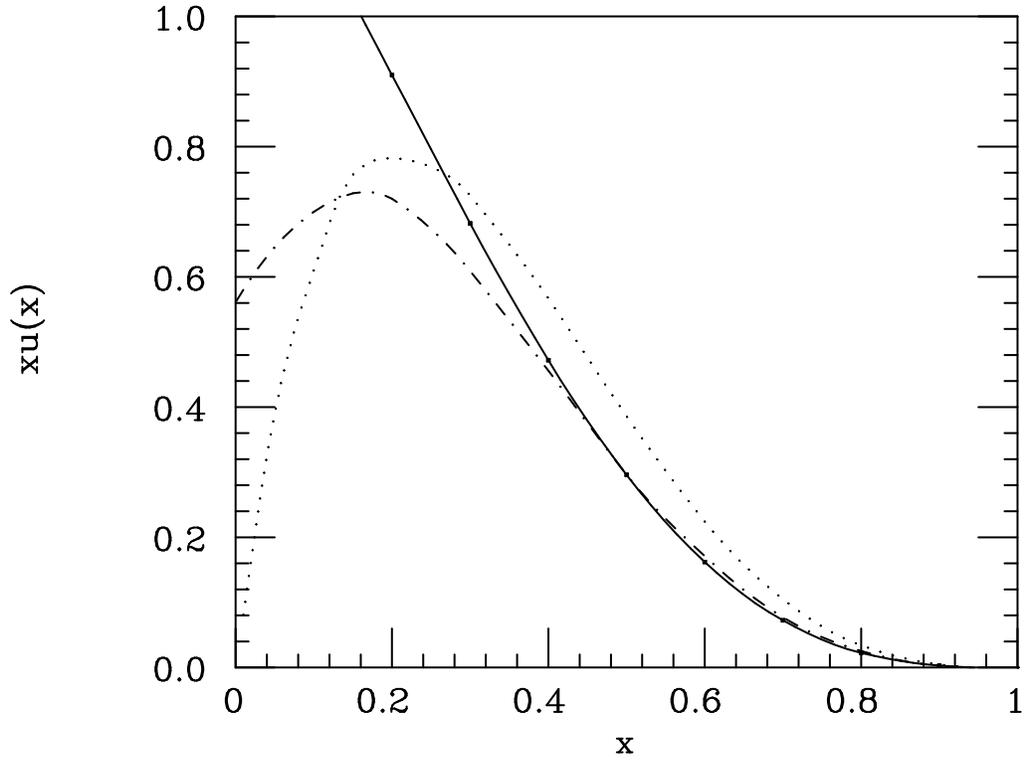,width=4in}}
\vglue 1in
\caption{Quark distribution $xu(x)$ of 5th column of Table 1 (dotted), Regge 
case of last column of Table 1 (solid) compared with BBS [12] (dot-dashed).}
\end{figure}

\begin{figure}[h]
\vglue 2in
\centerline{\psfig{figure=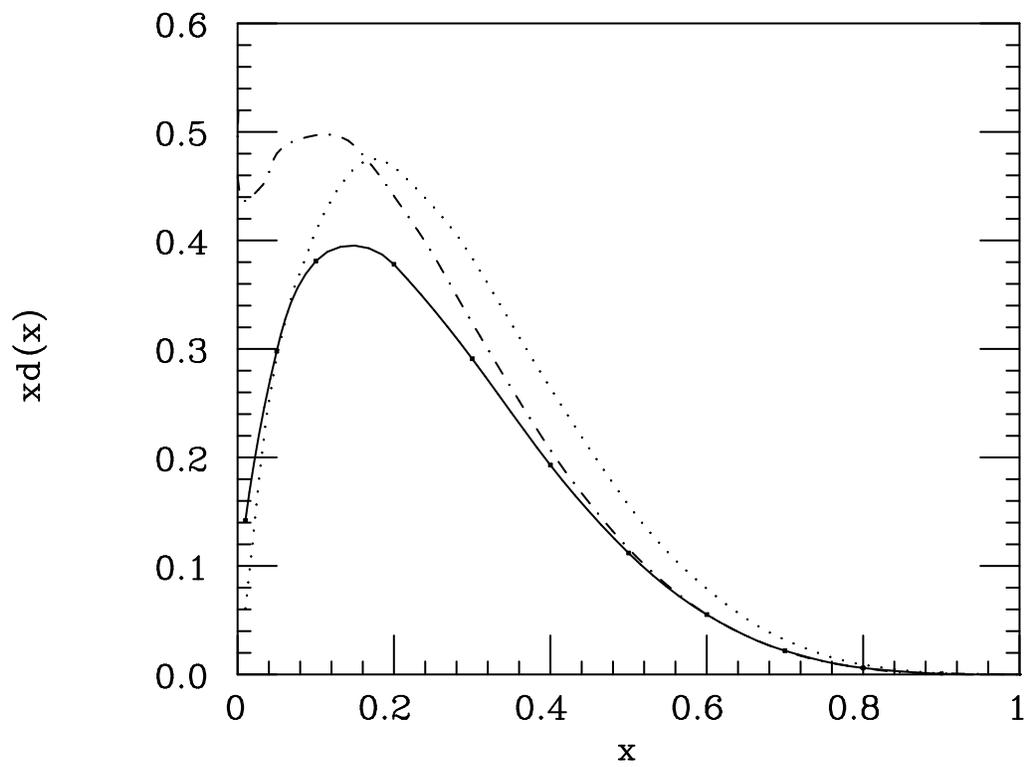,width=4in}}
\vglue 1in
\caption{Same as Fig.3 for down quark $xd(x)$.}
\end{figure}

\begin{figure}[h]
\vglue 2in
\centerline{\psfig{figure=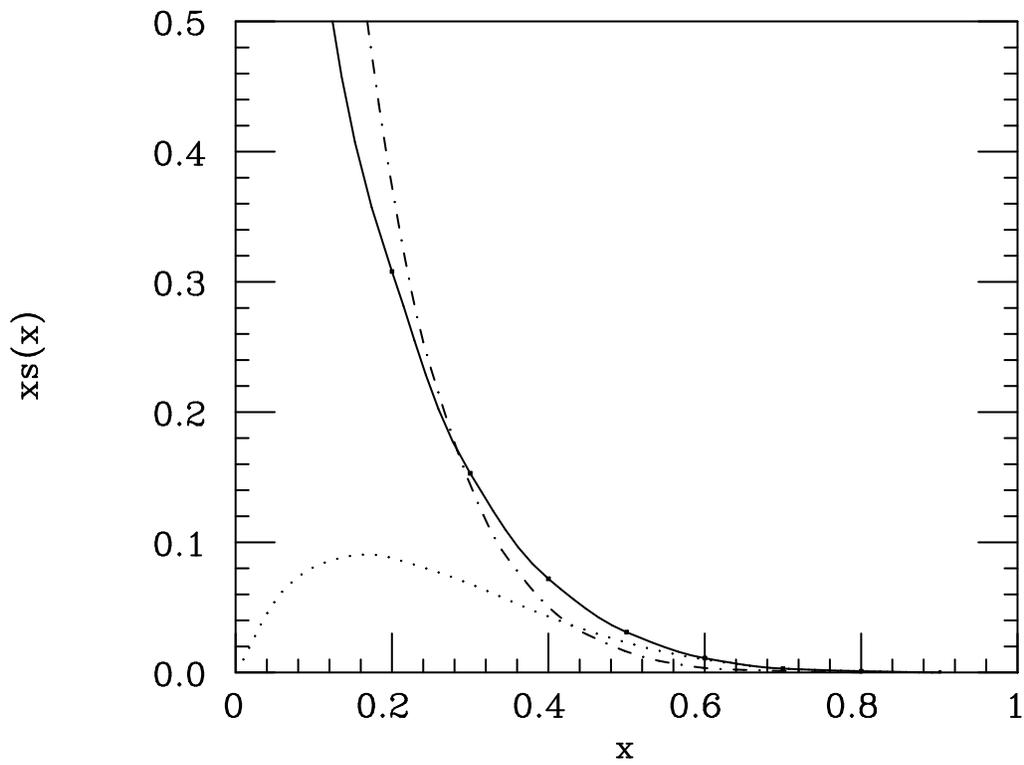,width=4in}}
\vglue 1in
\caption{Same as Fig.3 for strange quark $xs(x)$.} 
\end{figure}

\begin{figure}[h]
\vglue 2in
\centerline{\psfig{figure=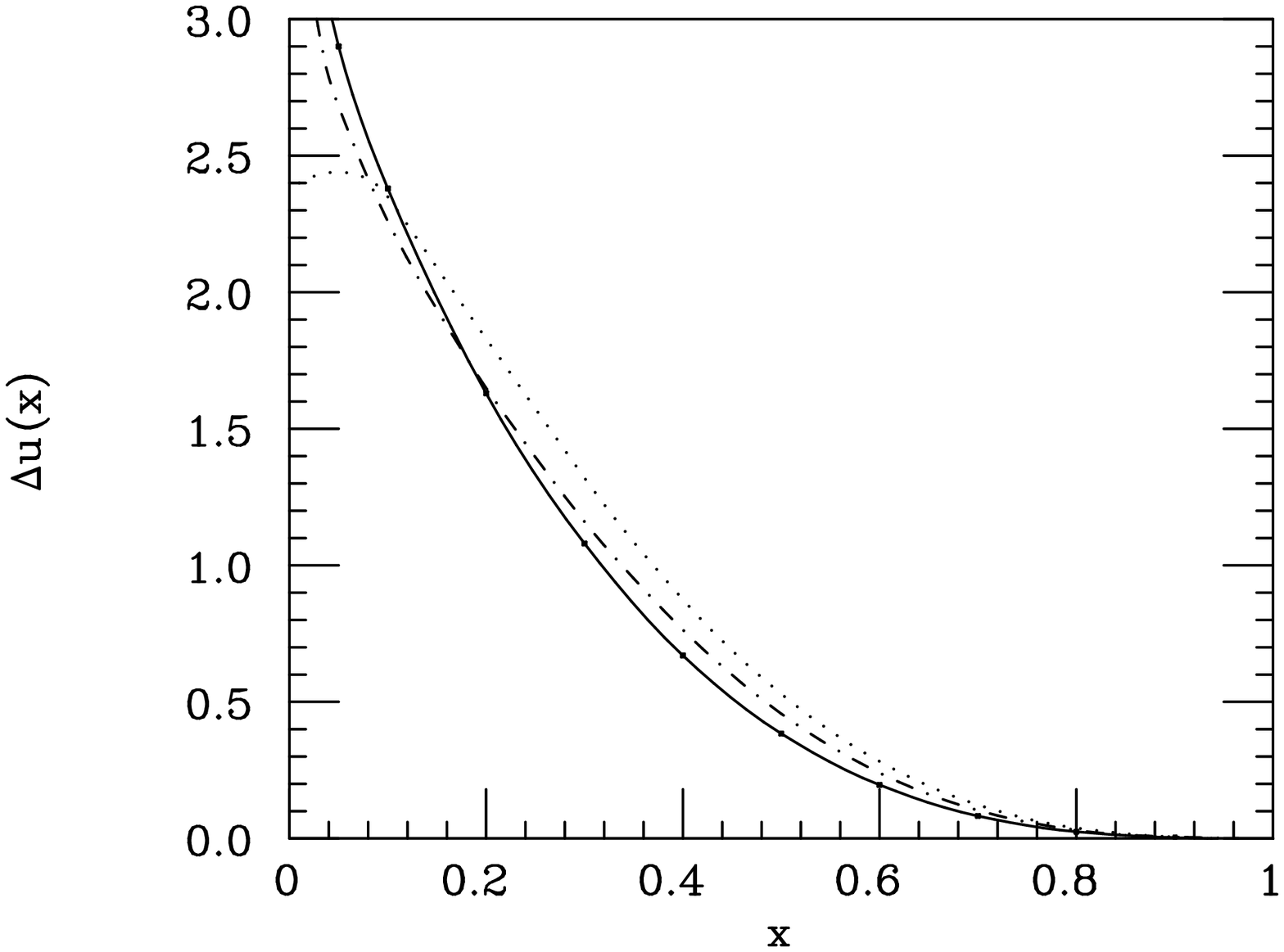,width=4in}}
\vglue 1in
\caption{Same as Fig.3 for polarized $\Delta u(x)$.} 
\end{figure}

\begin{figure}[h]
\vglue 2in
\centerline{\psfig{figure=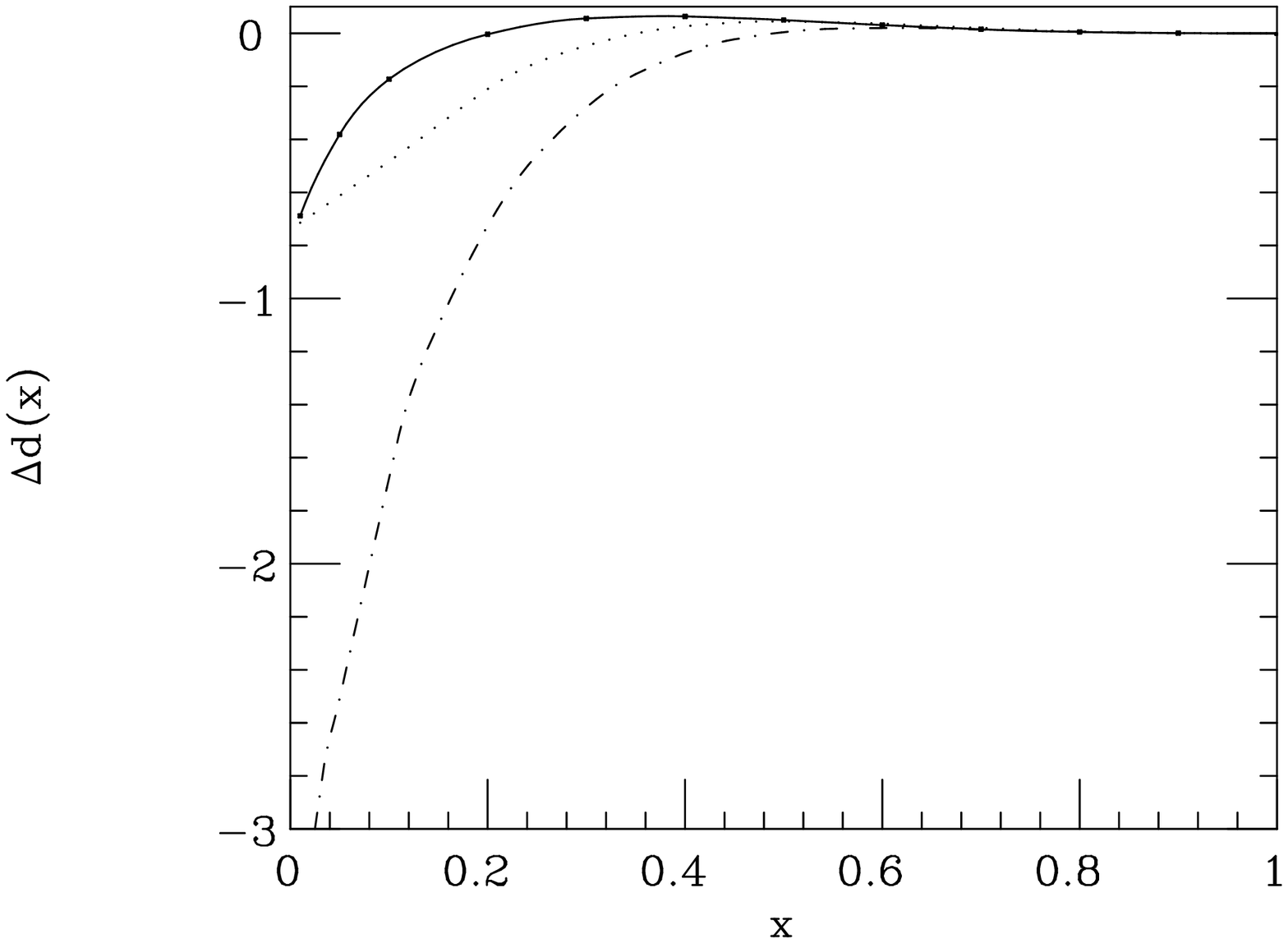,width=4in}}
\vglue 1in
\caption{Same as Fig.3 for polarized $\Delta d(x)$.}
\end{figure}

\begin{figure}[h]
\vglue 2in
\centerline{\psfig{figure=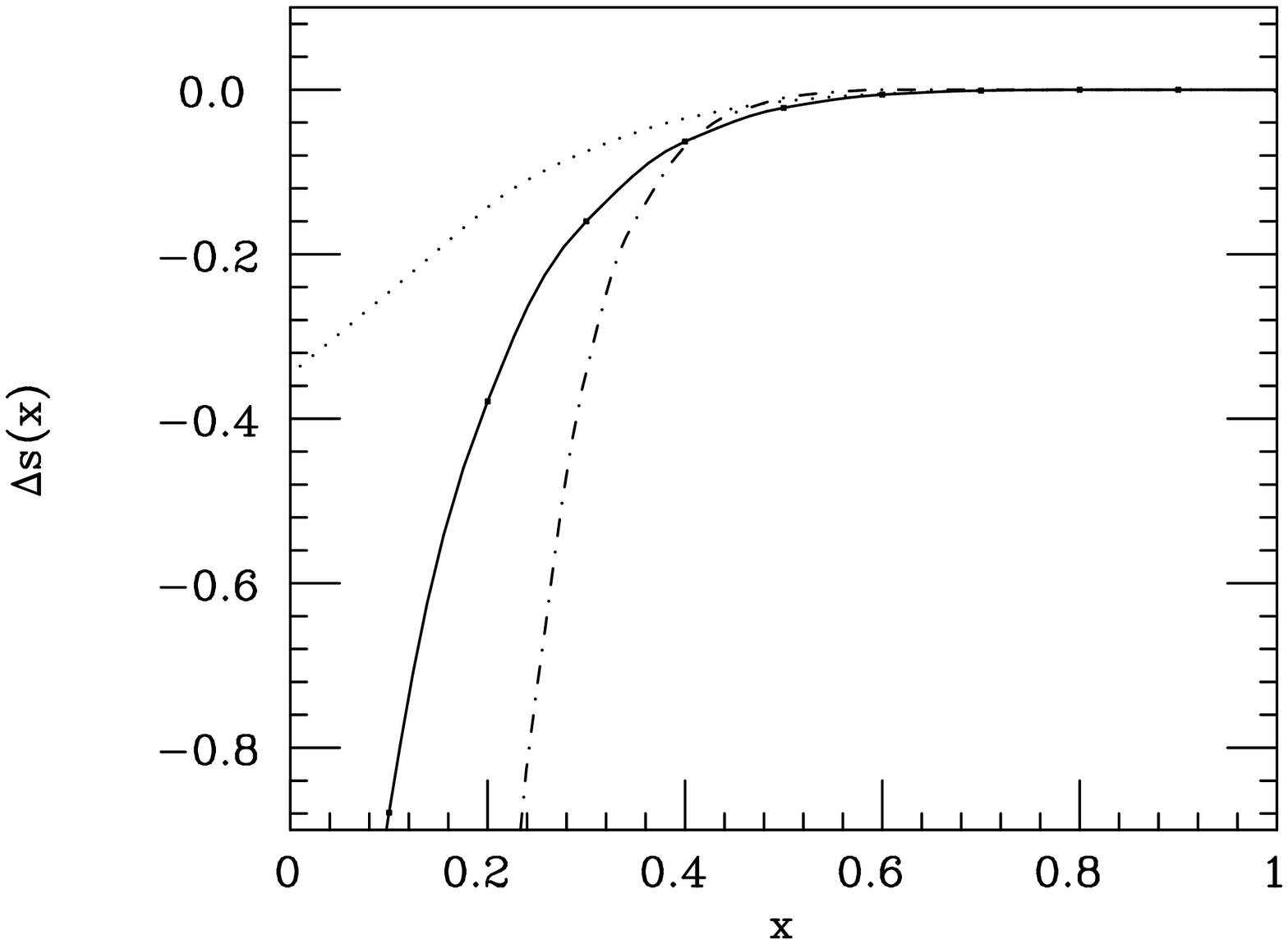,width=4in}}
\vglue 1in
\caption{Same as Fig.3 for polarized $\Delta s(x)$.}
\end{figure}

\begin{figure}[h]
\vglue 2in
\centerline{\psfig{figure=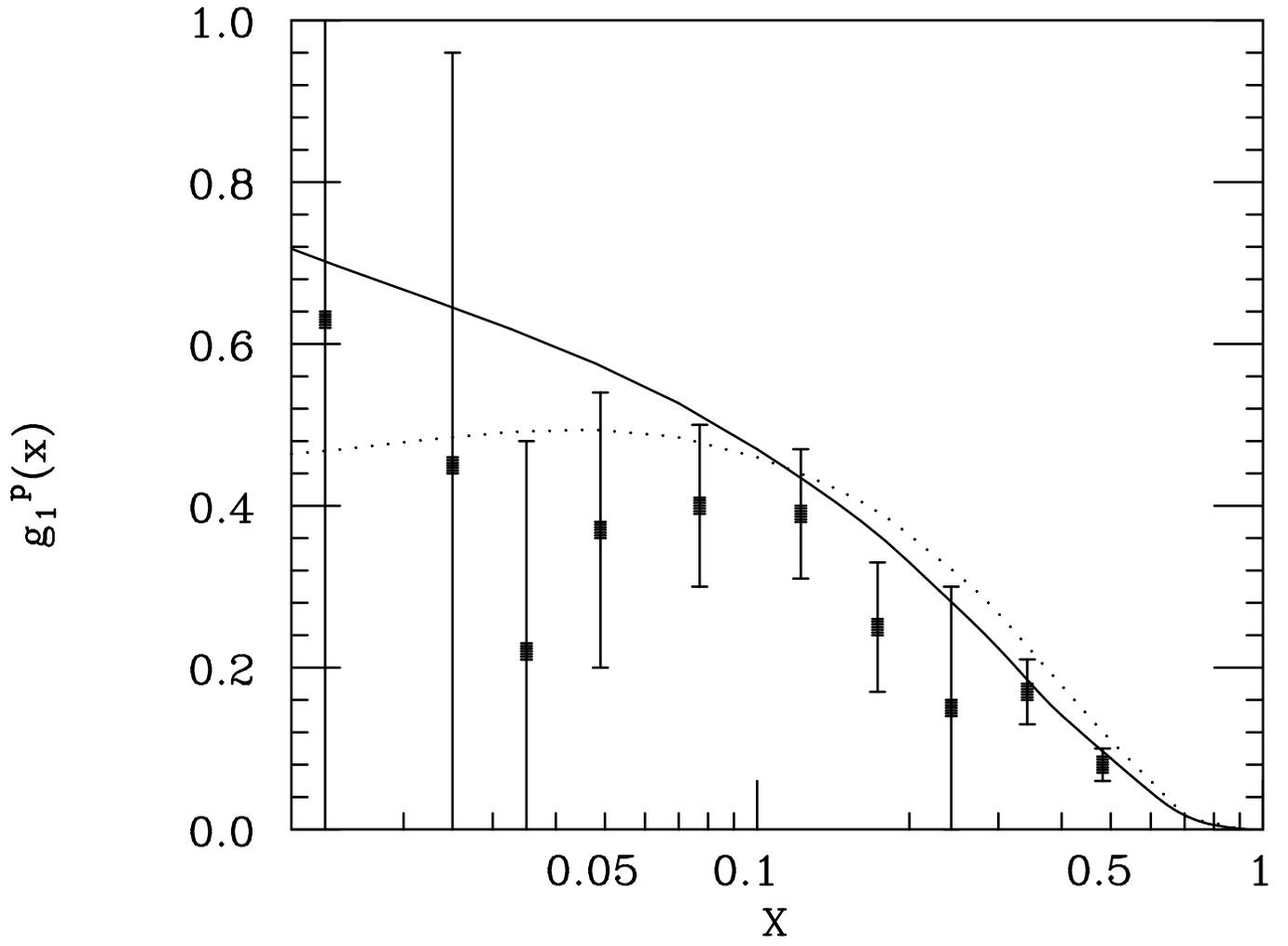,width=4in}}
\vglue 1in
\caption{Polarized structure function $g^{p}_{1}(x)$ of the proton; 
notation as in Fig.3. Data are from SMC, hep-ex/9702005.} 
\end{figure}

\begin{figure}[h]
\vglue 2in
\centerline{\psfig{figure=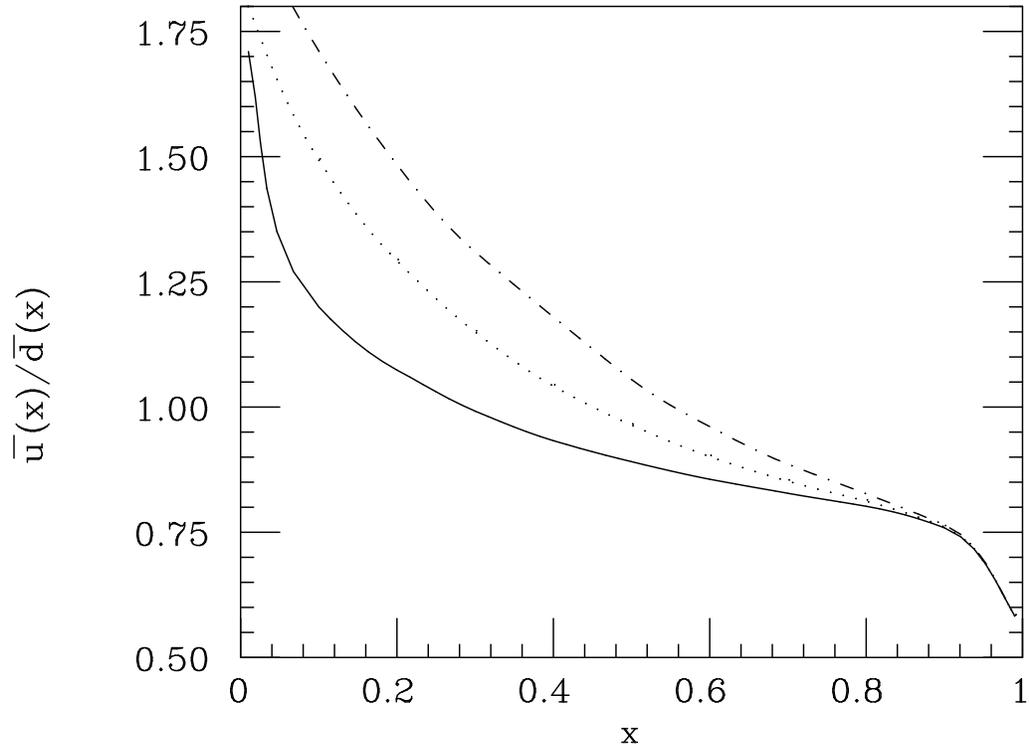,width=4in}}
\vglue 1in
\caption{Same as Fig.3 for the antiquark ratio $\bar u(x)/\bar d(x)$ } 
\end{figure}

\end{document}